\documentclass[11pt]{article}
\usepackage{latexsym}
\usepackage{epsfig}
\usepackage{graphicx}

\newcommand{\reals}{\mbox{${\rm I\!R }$}}
\newcommand{\nats}{\mbox{${\rm I\!N }$}}

\def\beq{\begin{eqnarray}}
\def\eeq{\end{eqnarray}}
\newcommand{\nn}{\nonumber}

\newcommand{\sujnu}{\sum_{j=0}^{\infty}}
\newcommand{\sulnu}{\sum_{l=0}^{\infty}}
\newcommand{\suani}{\sum_{b=0}^i}

\newcommand{\slnu}{\sum_{l=0}^\infty}

\newcommand{\ikma}{\int\limits_{\gamma}\frac{dk}{2\pi i}\,\,(k^2+m^2)^{-s}\frac{\partial}{\partial k}}
\newcommand{\fff}{\frac{\partial}{\partial z}}
\newcommand{\ddd}{\int\limits_{ma/\nu}^{\infty}dz\,\,}
\newcommand{\eee}{\left[\left(\frac{z\nu} a\right)^2-m^2\right]^{-s}}
\newcommand{\iinma}{\ddd\eee\fff}

\newcommand{\Res}{{\rm Res\,\,}}

\newcommand{\pa}{\partial}

\newcommand{\amed}{A_{-1}^{\nu}(s)}
\newcommand{\anud}{A_{0}^{\nu}(s)}
\newcommand{\aid}{A_{i}^{\nu}(s)}

\newcommand{\zend}{\zeta^{\nu}}
\newcommand{\zb}{\zeta_{{\cal N}}}

\newcommand{\zh}{\zeta_H}

\newcommand{\zba}{\zeta_{{\cal B}}}
\newcommand{\zeb}{\zeta_{{\cal B}}}

\newcommand{\pold}{D^{(d-1)}}
\newcommand{\hem}{a}

\newcommand{\facb}{\frac{(4\pi)^{d/2}}{|S^d|}}
\newcommand{\fac}{\frac{(4\pi)^{D/2}}{|S^d|}}

\newcommand{\sip}{\frac{\sin (\pi s)}{\pi}}

\newcommand{\mzs}{m^{-2s}}
\newcommand{\g}{\Gamma\left(}
\newcommand{\numr}{\left(\frac{\nu}{ma}\right)^2}
\newcommand{\rzs}{a^{2s}}
\newcommand{\lll}{\frac{(-1)^j}{j!}}
\newcommand{\ent}{d (\nu )}

\newcommand{\bou}{\cab_S^\mp}

\newcommand{\cab}{{\cal B}}
\newcommand{\cac}{{\cal C}}
\newcommand{\can}{{\cal N}}
\newcommand{\cam}{{\cal M}}
\newcommand{\cao}{{\cal O}}
\newcommand{\caz}{{\cal Z}}

\newcommand{\al}{\alpha}

\newcommand{\de}{\delta}
\newcommand{\ep}{\epsilon}

\newcommand{\la}{\lambda}
\newcommand{\om}{\omega}
\newcommand{\ze}{\zeta}

\newcommand{\De}{\Delta}
\newcommand{\Om}{\Omega}

\newcommand{\G}{\Gamma}

\newcommand{\bc}{boundary condition }
\newcommand{\bcs}{boundary conditions }
\newcommand{\zf}{zeta function }
\newcommand{\zfs}{zeta functions }

\newcommand{\ph}{\phantom}

\newcommand{\hks}{heat-kernel coefficients }

\renewcommand{\theequation}{\arabic{section}.\arabic{equation}}

\begin{document}

\title{Spectral functions in mathematics and physics}
\author{Klaus Kirsten\thanks{e-mail:Klaus.Kirsten@itp.uni-leipzig.de\newline
Present address: Department of Physics and Astronomy, The University
of Manchester, Oxford Road, Manchester UK M139PL}\\
Universit\"at Leipzig\\
Fakult\"at f\"ur Physik und Geowissenschaften\\
Institut f\"ur Theoretische Physik\\
Augustusplatz 10/11\\
04109 Leipzig\\
Germany}
\maketitle

\begin{abstract}

Spectral functions relevant in the context of quantum field theory
under the influence of spherically symmetric external conditions
are analysed. Examples comprise heat-kernels, determinants and
spectral sums needed for the analysis of Casimir energies. First,
we summarize that a convenient way of handling them is to use the
associated zeta function. A way to determine all its needed
properties is derived. Using the connection with the mentioned
spectral functions, we provide: i.) a method for the calculation
of heat-kernel coefficients of Laplace-like operators on
Riemannian manifolds with smooth boundaries and ii.) an analysis
of vacuum energies in the presence of spherically symmetric
boundaries and external background potentials.

\end{abstract}

\section{Introduction}
\setcounter{equation}{0} There is a notorious appearance of
spectral functions in many branches of mathematics and physics.
These functions are associated with suitable sequences of numbers
$\{\lambda_k\}_{k\in\nats}$, which for most applications are
eigenvalues of certain interesting, in most applications
Laplace-like, operators. A rich source of problems where spectral
functions are encountered is quantum field theory under the
influence of "external conditions". External means that the
condition is assumed to be known (as a function of space and time)
and only appears in the equation of motion of other fields, which
are to be quantized under these conditions. Given that the focus
of interest is on the influence of the external conditions, these
other fields are often assumed to be non-selfinteracting. It is in
this setting that the present work is to be viewed.

Under the described circumstances the action of the considered
theory will be quadratic in the quantized field. Using a path
integral formulation to describe the underlying quantized theory,
one encounters Gaussian functional integrals which lead to
functional determinants formally defined as $\prod_{k=1}^\infty
\lambda_k$. Making sense out of this kind of expressions is one of
the basic themes of quantum field theory. The present contribution
deals with scalar fields and will develop and apply techniques to
analyse determinants arising when the external conditions are
described by boundary conditions (Casimir effect) or when external
scalar background fields are present. We restrict to scalar fields
because the application of the provided ideas to spinor fields or
electromagnetic fields is then immediate (see
f.e.~\cite{bord9812060,bord9811015}).

In some more detail the situation considered will be described
by the action (we assume an Euclidean formulation)
\beq
S[\Phi ] &=& -\frac 1 2 \int_\cam d x \Phi (x) (\Box_E
  -V(x) )  \Phi (x) , \label{eq1.1a}
\eeq
for a scalar field $\Phi$ in the background potential $V(x)$.
Here, $\cam$ is a $D$-dimensional Riemannian manifold and $dx$
its volume element.

The corresponding field equation is
\beq
(\Box_E -V(x) ) \Phi (x) &=& 0 .\label{eq1.1c}
\eeq

If boundaries are present, the equations of motion are
supplemented by boundary conditions to be specified later.

Physical properties of the systems are conveniently described by means of the
path-integral functionals (an infinite normalization constant is neglected)
\beq
\caz [V] &=& \int D\Phi e^{-S[\Phi ]} ,\label{eq1.1e}
\eeq
where the functionals are taken over all fields satisfying,
if applicable, the boundary conditions imposed.

Under the circumstances described, the effective action
can (at least formally) easily be computed to be (assume that
there are no zero modes; if these are present they have to be omitted
because otherwise the determinant is trivially zero),
\beq
\G [V]= -\ln \caz [V] &=& \frac 1 2 \ln \mbox{det} \left[
         ( -\Box_E +V(x) )/ \mu^2 \right] .\nn\eeq

Here, $\mu$ is an arbitrary parameter with dimension of a mass, to
make the argument of the logarithm dimensionless. The operators
involved are thus Laplace-like operators, extensively dealt with
afterwards. Let us write them in the unified form also used later,
namely
\beq P = -g^{\rho \nu} \nabla_\rho \nabla _\nu -E
,\label{eq1.1g} \eeq
where $g^{\rho \nu}$ is the Riemannian metric
of the manifold $\cam$, $\nabla$ is a connection and $E$ an
endomorphism defined on $\cam$. We are thus confronted with the
task of calculating expressions of the type
\beq \Gamma [V] =\frac
1 2 \ln [\mbox{det} (P / \mu^2)] .\label{eq1.2} \eeq

Clearly, expression (\ref{eq1.2}) is not defined because the
eigenvalues $\lambda_n$ of $P$, \beq P\phi_n = \lambda_n \phi_n,
\label{eq1.3} \eeq grow without bound for $n\to \infty$. Of
course, there are various possible regularization procedures; let
us mention only Pauli-Villars, dimensional regularization and zeta
function regularization. Here, we will use \zf regularization
\cite{dowk76-13-3224,hawk77-55-133} because it is mathematically
appealing as well as (probably) the most convenient one in the
context of our work. The basic idea is to generalize the identity,
valid for a $(N\times N)$-matrix $P$,
\beq \ln \mbox{det}\,\, P =
\sum_{n=1}^N \ln \lambda_n = - \frac d {ds}  \sum_{n=1}^N
\lambda_n^{-s} |_{s=0} = -\frac d {ds}
         \zeta_P (s) |_{s=0} , \nn
\eeq with the zeta function \beq \zeta_P (s) = \sum_{n=1}^N
\lambda_n^{-s} ,\nn \eeq to the differential operator $P$
appearing in (\ref{eq1.3}) by \beq \ln \mbox{det}\,\, P =
-\zeta_P' (0) , \label{eq1.4} \eeq with
\beq \zeta_P (s) =
\sum_{n=1}^\infty \lambda_n^{-s} .\label{eq1.5} \eeq

That this definition is in fact sensible is a result of deep
mathematical theorems on the analytical structure of $\zeta_P (s)$
(explained in the following).

First of all, due to a classical theorem of Weyl \cite{weyl12-71-441}, which
says that for a second order elliptic differential operator the eigenvalues
behave asymptotically for $n\to\infty$ as
\beq
\lambda_n^{D/2} \sim \frac{2^{D-1} \pi^{D/2} D \G (D/2)}
                {\mbox{vol}(\cam)} n,\nn
\eeq the representation (\ref{eq1.5}) of $\zeta_P (s)$ is valid
for $\Re s > D/2$. In order to use definition (\ref{eq1.4}), the
question arises of how to analytically continue $\zeta_P (s)$ to
the left and to determine its analytic structure. This is very
elegantly done by using an integral representation of the
$\Gamma$-function to write (see f.e. \cite{voro87-110-439}), still
for $\Re s > D/2$, \beq \zeta_P (s) = \frac 1 {\G (s)}
\int_0^\infty t^{s-1} K(t) , \label{eq1.6} \eeq with the
heat-kernel \beq K(t) = \sum_{n=1}^\infty e^{-\lambda_n t} .\nn
\eeq

For $t\to\infty$ the integral is well behaved due to the
exponential damping coming from $K(t)$ (for simplicity we assume a
positive definite operator $P$). Possible residues only arise from
the $t\to 0$ behaviour of the integrand, and we need information
on $K(t)$ for $t\to 0$. This is given by the heat-kernel expansion
\cite{mina49-1-242,mina53-17-158,gilk95b} \beq K(t) \sim
\sum_{l=0,1/2,1,...}^\infty a_l (P) t^{l-D/2} ,\label{eq1.6a} \eeq
with the heat-kernel coefficients $a_l (P)$ depending, of course,
explicitly on the operator $P$ together with the boundary
conditions chosen. If the manifold has no boundary, the
coefficients with half-integer index vanish. Splitting the
integral for example into $\int_0^1 dt + \int_1^\infty dt$ the
following connection is obtained \cite{seel67b}, \beq \Res
(\zeta_P (s) \G (s) ) |_{s=D/2-l } = a_l (P) , \label{eq1.7} \eeq
or, showing the information contained more clearly, for
$z=D/2,(D-1)/2,...,1/2,-(2n+1)/2, n\in\nats_0$, \beq \Res \zeta_P
(z) = \frac{ a_{D/2-z}(P)}{\G (z)} , \label{eq1.8} \eeq and for
$q\in\nats_0$, \beq \zeta_P (-q) = (-1)^q q! a_{D/2 +q}
(P).\label{eq1.9} \eeq

This clearly shows that, for manifolds without boundaries, the
poles are located at $z=D/2,D/2-1,...,1$ for $D$ even, and
$z=D/2,D/2-1,...,1/2, -(2n+1)/2$, $n\in\nats_0$, for $D$ odd. In
addition, for $D$ odd and $q\in\nats_0$, one gets $\zeta_P (-q)
=0$. For manifolds with boundary, additional possible poles appear
and here one finds poles at $D/2,(D-1)/2,...,1/2,-(2n+1)/2$,
$n\in\nats_0$. In all cases, $\zeta_P (s)$ is an analytical
function in a neighbourhood of $s=0$ such that eq. (\ref{eq1.4})
is well defined. This definition was first used by the
mathematicians Ray and Singer \cite{ray71-7-145}, when trying to
give a definition of the Reidemeister-Franz torsion
\cite{fran35-173-245}.

Let us stress that the expansion (\ref{eq1.6a}) of the heat-kernel
and the connection (\ref{eq1.7}) with the zeta function strongly
depend on the assumptions made, and strictly they hold only if we
are dealing with a {\it second order elliptic} differential
operator on a {\it smooth compact Riemannian} manifold with a {\it
smooth} boundary and {\it local elliptic} boundary conditions
\cite{gilk95b}. For example, in the context of admissible pseudo
differential operators and global spectral boundary conditions,
different powers of $t$ can be involved in (\ref{eq1.6a}) and, in
addition, ($\ln t$)-terms might appear. This was noted starting
with \cite{duis75-29-39} and later, for example, in
\cite{brun85-58-133,grub86b,kuro88-64-21}, where the asymptotic
expansions for these cases can be found. That indeed generically
all possible terms are present (what means generically can be made
precise) has been shown recently \cite{gilk98-23-777}. As a
consequence, the simple poles of the zeta function may well be
located at other points depending on the power of $t$ appearing in
the asymptotic small-$t$ expansion of $K(t)$, and, if there are
$\ln t$-terms, double poles may be present as well. Also in this
more general context, analogous relations to equation
(\ref{eq1.7}) may be stated \cite{jorg93b}, but here we will make
no use of these and, therefore, we won't bother to state more
details.

A case of particular interest is the manifold $\cam = S^1 \times
M_{s}$ with the operator \beq P=-\frac{\partial^2}{\pa \tau^2}
+P_s ,\label{eq1.10} \eeq and $P_s$ Laplace-like. Imposing
periodic \bcs in the $\tau$-variable this is finite temperature
quantum field theory for a scalar field and the perimeter $\beta$
of the circle plays the role of the inverse temperature. Under the
assumption that the potential $V$ and the \bcs are static, $P_s$
is a purely spatial operator and depends only on coordinates on
$M_s$. In this context, it is known that the energy of the system
is by definition
\beq E= - \frac \pa {\pa \beta} \ln {\cal Z} =
-\frac 1 2
   \frac \pa {\pa \beta}    \zeta_{P/\mu^2 } ' (0) ,\nn
\eeq
and it seems natural to use \cite{cogn92-33-222}
\beq
E_{vac} = \lim_{\beta \to\infty} E = \frac 1 2 FP\,\,\zeta_{P_s} (-1/2)
      -\frac 1 {2\sqrt{4\pi}} a_{D/2}(P_s) \ln \tilde \mu^2 , \label{eq1.13}
\eeq
with the scale $\tilde \mu = (\mu e / 2)$ as the definition of
the vacuum energy. Here, $FP$ means finite part for cases where
$\zeta_{P_s} (-1/2)$ has a pole. Due to the presence of the
arbitrary scale $\mu$, it is seen that the vacuum energy is
ambiguous, which generally causes problems to achieve a physically
sensible answer. However, this is the way the Casimir energy is
usually defined (see for example
\cite{dowk78-11-895,ambj83-147-1,dowk84-1-359,dola92-148-139,caru91-43-1300,cogn92-33-222,byts96-266-1,eliz94b,eliz95b,blau88-310-163,albu97-55-7754,bene9809081}),
the idea for the derivation going back already to Gibbons
\cite{gibb77-60-385}. The ambiguity may be discussed within
renormalization group equations found by demanding
\cite{wies94-233-125,cogn93-48-790} \beq \mu \frac d {d\mu} \G [V]
=0 .\nn \eeq

Eq.~(\ref{eq1.13}) then clearly shows that the total energy of the
system must contain all terms present in $a_{D/2} (P_s)$ in order
to define the needed counterterms and running coupling constants.
These terms describe the energy of the external fields or are the
energy needed to get a model for the boundary conditions. If
$a_{D/2} (P_s)\neq 0$, the Casimir energy is determined only up to
terms proportional to $a_{D/2} (P_s)$ and this finite ambiguity
can (in principle) only be eliminated by experiments
\cite{blau88-310-163}. A possible way to fix the ambiguity for
massive fields will be discussed in Section \ref{sec6}. If, on the
contrary, $a_{D/2} (P_s) =0$, eq.~(\ref{eq1.13}) gives a unique
answer for the energy.

Eq.~(\ref{eq1.13}) is also the definition one would naively use.
To see this, write the Hamilton operator formally as \beq H =
\sum_k E_k \left(N_k +\frac 1 2 \right) , \nn \eeq
with $N_k$ the
number operator, to obtain for the vacuum energy
\beq E_{vac} =
<0|H|0>=\frac 1 2 \sum_k E_k .\label{casimir} \eeq

The regularization \beq E_{vac} &=& \frac {\mu^{2s}} 2 \sum_k
(E_k^2)^{1/2-s} |_{s=0}
     = \frac{\mu^{2s}} 2 \zeta _{P_s} (s-1/2) |_{s=0} \nn\\
 &=& \frac 1 2 FP\,\,\zeta _{P_s} (-1/2) +
\frac 1 2 \left( \frac 1 s + \ln \mu^2 \right) \Res
\zeta _{P_s} (-1/2) \nn\\
&=& \frac 1 2 FP\,\,\zeta _{P_s} (-1/2)
- \left( \frac 1 s + \ln \mu^2 \right) \frac 1 {2\sqrt{4\pi}} a_{D/2}(P_s)
\label{eq1.13a}
\eeq
is clearly equivalent to eq. (\ref{eq1.13}), the only difference being that
renormalization now involves infinities.

Let us stress that, although we have focused on
non-selfinteracting scalar fields, the calculation of determinants
like in eq.~(\ref{eq1.2}) is of great relevance in several areas
of modern theoretical physics. In many situations the potential is
provided by classical solutions to nonlinear field equations.
Quantizing a theory about these solutions leads to the same kind
of determinants as discussed above \cite{raja82b}. The classical
solutions involved may be monopoles
\cite{thoo74-79-276,poly74-20-194}, sphalerons
\cite{klin84-30-2212} or electroweak Skyrmions
\cite{gips81-183-524,gips84-231-365,ambj85-256-434,eila86-56-1331,frie77-15-1694,
frie77-16-1096,skyr61-260-127,skyr72-31-556,adki93-228-552}. The
determinant in these external fields enters semiclassical
transition rates as well as the nucleation of bubbles or droplets
\cite{cole77-15-2929,call77-16-1762}. In general, the classical
fields are inhomogeneous configurations and, as a rule, the
effective potential approximation to the effective action, where
quantum fluctuations are integrated out about a constant classical
field, is not expected to be adequate. The derivative expansion
\cite{chan85-54-1222} improves on this by accounting for spatially
varying background fields. Being a perturbative approximation it
has however its own limitations. Having in mind that even the
classical solutions are often known only numerically, it is clear
that it is desirable to have a numerical procedure to determine
the quantum corrections. Some contributions in this direction are
\cite{baac90-47-263,baac93-48-5648,lee94-49-4101,brah94-49-4094}.

To summarize, as we have briefly described, the most relevant
spectral functions in quantum field theory under external
conditions are determinants, eq.~(\ref{eq1.2}), respectively the
spectral sum, eq.~(\ref{casimir}), and the heat-kernel,
eq.~(\ref{eq1.6a}), describing the ambiguity in Casimir and ground
state energy calculations. It is the aim of the present work, to
provide and apply techniques for the analysis of all these
spectral functions. As we have seen in the general introduction
above, it is the zeta function associated with the spectrum which
can be used as the organizing quantity for all the calculations to
be done.

The starting point of our investigation was the aim to develop a
machinery adequate to deal with spherically symmetric external
conditions, the motivation being that many of the classical
solutions have this property. In all spherically symmetric
problems, the total angular momentum is a conserved quantity and,
as a result, eigenvalues will be labelled by this angular momentum
and an additional main quantum number. Compared to one-dimensional
problems, where many calculations can be done exactly (see
f.e.~\cite{wipf85-58-531,boch92-46-5550,bord95-28-755,nast9802074}),
one encounters here the common technical complication of angular
momentum sums. Any spherically symmetric example can help to
understand the difficulties arising therefrom.

Intuitively, to impose boundary conditions on a spherical shell
seems simpler than dealing with an arbitrary spherically symmetric
external field, because eigenfunctions are known explicitly.
However, let us stress that analytical expressions for the
eigenvalues are not available which complicates the analysis
considerably. Based on the knowledge of the eigenfunctions, an
important achievement of our consideration is that various
properties of spectral functions associated with a spectrum which
is not known explicitly can be determined. We explain this
procedure in detail for the spectrum of the Laplace operator on
the three dimensional ball \cite{bord96-37-895}.

Having a very good (if not complete) knowledge of the zeta
function at hand, and using the connections explained above, the
obtained results are applied to the calculation of the associated
heat-kernel coefficients and Casimir energies. The applications
are grouped according to their complexity: i.) the heat-kernel
coefficients, which are "local" quantities and whose determination
is purely analytical, ii.) Casimir energies, which are non-local
and where additional numerical work is needed.

For the mentioned reason we start, in Section 3, with the
consideration of heat-kernel coefficients. An extremely important
aspect of this calculation is that it is {\it not} just a special
case calculation, but very rich information about \hks for
Laplace-like operators on arbitrary compact Riemannian manifolds
evolves. Supplemented by other (already known) techniques
involving conformal variations and the application of index
theorems \cite{gilk95b}, it allowed for the determination of new
coefficients for all types of boundary conditions
\cite{bord96-182-371,dowk99-,dowk9803094,dowk9806168,dowk98pro}
(including Dirichlet, Robin, mixed, oblique and spectral boundary
conditions).

In Section \ref{sec6} we consider Casimir energies in the presence
of spherically symmetric boundaries. Electromagnetic field
fluctuations in a spherical shell were considered already a long
time ago in the context of a simple model for an electron
\cite{casi53-19-846,boye68-174-1764,milt78-115-388,bali78-112-165},
where the zero-point energy of the electromagnetic field was
supposed to stabilize the classical electron. However, as was
first shown by Boyer \cite{boye68-174-1764}, a repulsive force is
the result and the simple model fails.

Our focus here will be on the influence of a mass $m$ of the
quantum field, and we will consider a scalar field with Dirichlet
boundary conditions. Based on the only previously  existing
complete calculation for planar boundaries at a distance $R$
\cite{plun86-134-87}, the general believe is that for $mR\gg 1$
the Casimir energy is exponentially small and thus of very short
range. As we will see and explain, this is due to the planar
boundaries and does not hold if the boundaries are curved. It
might even happen that the Casimir energy changes sign as a
function of the mass. This surprising result justifies to stand
the additional complication due to the non-vanishing mass.
Although we restrict for simplicity to scalar fields with
Dirichlet boundary conditions, higher spin fields with various
boundary conditions as f.e.~MIT bag-boundary conditions can be
treated as well along these lines (see f.e.
\cite{fran97-55-2477,fran94-50-2908,eliz98-31-1743}).

Afterwards, we analyze the influence of external background fields
\cite{bord96-53-5753,bord99u}. Although, in general, not even the
eigenfunctions are explicitly given, this knowledge is replaced by
the information available from scattering theory. So we will
express the ground state energy by the Jost function, which is
known (at least) numerically by solving the Lippmann-Schwinger
equation. Applying the formalism developed in Section \ref{sec3.1}
for the treatment of the angular momentum sums, quite explicit
results for the ground state energies are obtained. An example
shows the typical features of the dependence of the ground state
energy on the external potential.

Section \ref{sec3.1} provides the basis for the rest of this work.
However, the sections on the determination of heat-kernel coefficients and
the calculation of Casimir and ground state energies are completely independent
and can be read separately.

\section{Developing the formalism: Scalar field on the three dimensional ball}
\setcounter{equation}{0}
\label{sec3.1}

To start we focus our interest on the zeta function of the operator
$(-\Delta +m^2)$ on the three dimensional ball
$B^3=\{x\in\reals ^3 ; |x|\leq a\}$ endowed with Dirichlet
boundary conditions. The zeta function is formally defined as
\beq
\zeta (s) =\sum_k \lambda_k ^{-s},\label{eq2.1}
\eeq
with the eigenvalues $\lambda_k$ being determined through
\beq
(-\Delta +m^2) \phi_k (x) =\lambda_k \phi _k (x)\label{eq2.2}
\eeq
($k$ is in general a multiindex here), together with Dirichlet boundary
conditions.
It is convenient to introduce a spherical coordinate basis, with $r=|x|$
and the angles $\Omega = (\theta, \varphi )$.
In these coordinates, a complete set of solutions of eq.~(\ref{eq2.2})
together with the boundary conditions may be given in
the form
\beq
\phi_{l,m,n}(r,\Omega )=r^{-1/2}J_{l+1/2}(w_{l,n}r)
Y_{lm}(\Omega ), \label{eq2.3}
\eeq
with the Bessel function $J_{l+1/2}$ and
$Y_{lm}(\Om )$ the spherical surface harmonics
\cite{erde55b}. The $w_{l,n}$
$(>0)$ are determined through the boundary condition by
 \beq
J_{l+1/2}(w_{l,n}a) =0 . \label{eq2.4} \eeq

In this notation, using $\lambda_{l,n}=w_{l,n}^2+m^2$, the zeta
function can be given in the form \beq \zeta (s)
=\sum_{n=0}^{\infty} \sulnu (2l+1) (w_{l,n}^2
+m^2)^{-s},\label{eq2.5} \eeq where  $w_{l,n}$ is defined as the
n-th root of the l-th equation (\ref{eq2.4}). Here, the sum over
$n$ is extended over all possible roots $w_{l,n}$ on the positive
real axis, and $(2l+1)$ is the number of independent harmonic
polynomials, which defines the degeneracy of each value of $l$ and
$n$ in three dimensions.

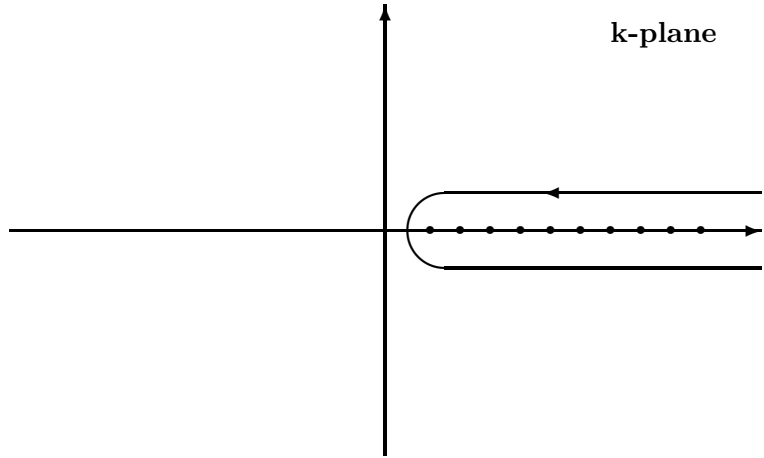
\begin{figure}[ht]
\setlength{\unitlength}{1cm}

\begin{center}
\begin{picture}(10,6.5)
\thicklines

\put(0,3){\vector(1,0){10}}
\put(5.0,0){\vector(0,1){6}}
\put(10.0,3){\oval(9.4,1)[l]}  \put(7.5,3.5){\vector(-1,0){.4}}
\multiput(5.6,3)(.4,0){10}{\circle*{.1}}

\put(8.0,5.5){{\bf k-plane}}

\end{picture}
\caption{\label{fig1}Contour $\gamma$}
\end{center}
\end{figure}

Eq.~(\ref{eq2.5}) may be written under the form of a contour
integral on the complex plane, \beq \zeta (s)= \sulnu (2l+1)
\ikma\ln J_{l+1/2}(ka),\label{eq2.6} \eeq where the contour
$\gamma$ runs counterclockwise and must enclose all the solutions
of (\ref{eq2.4}) on the positive real axis, see Fig.~\ref{fig1}
(for a similar treatment of the zeta function as a contour
integral see \cite{kame92-7-3713,barv92-219-201,bord95-28-755}).
Clearly, $(\pa / \pa k) \ln J_{l+1/2} (ka) = aJ_{l+1/2} ' (ka) /
J_{l+1/2} (ka)$, having simple poles at the solutions of
eq.~(\ref{eq2.4}) and, by the residue theorem, definition
(\ref{eq2.5}) is reproduced. This representation of the zeta
function in terms of a contour integral around some circuit
$\gamma$ on the complex plane, eq.~(\ref{eq2.6}), is the first
step of our procedure.

As it stands, the representation (\ref{eq2.6}) is valid for $\Re s
>3/2$. However, we are especially interested in the properties
of $\zeta(s)$ in
the range $\Re s < 3/2$ and
thus, we need to perform the analytical continuation to the left.
Before considering in detail the $l$-summation, we will first proceed
with the $k$-integral alone.

The first specific idea is to shift the integration contour and
place it along the imaginary axis. For $k\to 0$, to leading order,
one has the behaviour $J_\nu (k) \sim k^\nu/(2^\nu \G (\nu +1) )$
and, in order to avoid contributions coming from the origin $k=0$,
we will consider (with $\nu =l+1/2$) the expression \beq \zend (s)
= \ikma \ln \left(k^{-\nu} J_{\nu} (ka) \right),\label{eq2.7} \eeq
where the additional factor $k^{-\nu}$ in the logarithm does not
change the result, for no additional pole is enclosed. Using the
relations $J_\nu (ik) = e^{i\pi\nu} J_\nu (-ix)$ and $I_\nu (x) =
e^{-ix\pi /2} J_{\nu} (ix)$ \cite{grad65b}, one then easily
obtains \beq \zend  (s)=\sip \int\limits_m^{\infty}dk\,\,
[k^2-m^2]^{-s}\frac{\partial}{\partial k} \ln \left(k^{-\nu
}I_{\nu} (ka) \right)\label{eq2.8} \eeq valid in the strip $1/2
<\Re s <1$. Here, the upper restriction, $\Re s < 1$, is imposed
by the behaviour $(k-m)^{-s}$ of the integrand at the lower
integration bound $k\to m$. For $k\to\infty$ one uses instead
$I_\nu (k) \sim e^k/\sqrt{2\pi k}$ to find the behaviour $k^{-2s}$
and, thus, the restriction $1/2 < \Re s$. It gets clear that, with
$m=0$, this representation is defined for no value of $s$ and a
slightly more difficult procedure has to be used
\cite{eliz93-26-2409,lese94-27-2483}. We prefer to include the
mass $m$ and consider the limit $m\to 0$ (whenever needed) at a
stage of the calculation where the limit will be well defined.

Given that the interesting properties of the zeta function (namely
nearly all heat-kernel coefficients of $(-\Delta)$, the
determinant and the Casimir energy) are encoded at the left of the
strip $1/2 < \Re s < 1$, how can we find its analytical
continuation to the left? As explained, the restriction $1/2 < \Re
s$ is a result of the behaviour of the integrand for $k\to
\infty$. If we subtract this asymptotic behaviour from the
integrand in (\ref{eq2.8}), the strip of convergence will
certainly move to the left, and the hope will be that the
asymptotic terms alone can be treated analytically, whereas
$\zeta^\nu (s)$ with the asymptotic terms subtracted can be
treated numerically. In detail, one needs to make use of the
uniform asymptotic expansion of the Bessel function $I_{\nu} (k)$
for $\nu \to \infty$ as $z=k/\nu$ fixed \cite{abra70b} which, as
we will see, guarantees that integration as well as summation lead
to analytical expressions. One has \beq I_{\nu} (\nu z) \sim \frac
1 {\sqrt{2\pi \nu}}\frac{e^{\nu \eta}}{(1+z^2)^{\frac 1
4}}\left[1+\sum_{k=1}^{\infty} \frac{u_k (t)} {\nu
^k}\right],\label{eq2.9} \eeq with $t=1/\sqrt{1+z^2}$ and $\eta
=\sqrt{1+z^2}+\ln [z/(1+\sqrt{1+z^2})]$. The first few
coefficients are listed in \cite{abra70b}; higher coefficients are
immediate  to obtain by using the recursion \cite{abra70b} \beq
u_{k+1} (t) =\frac 1 2 t^2 (1-t^2) u'_k (t) +\frac 1 8
\int\limits_0^t d\tau\,\, (1-5\tau^2 ) u_k (\tau ),\label{eq2.10}
\eeq starting with $u_0 (t) =1$. As is clear, all the $u_k (t)$
are polynomials in $t$. Furthermore, the coefficients $D_n (t)$
defined by \beq \ln \left[1+\sum_{k=1}^{\infty} \frac{u_k (t)}{\nu
^k}\right] \sim \sum_{n=1}^{\infty} \frac{D_n (t)}{\nu ^n}
,\label{eq2.11} \eeq are easily found with the help of a simple
computer program. For example one has \beq D_1 (t) &=& \frac 1 8 t
-\frac 5 {24} t^3 , \nn\\ D_2 (t) &=& \frac 1 {16} t^2 -\frac 3 8
t^4 +\frac 5 {16} t^6.
 \label{eq2.11a}
\eeq

By adding and subtracting $N$ leading terms of the asymptotic
expansion, eq.~(\ref{eq2.11}), for $\nu \to \infty$,
eq.~(\ref{eq2.8}) may be split into the following pieces, \beq
\zend  (s) =Z^{\nu}(s) +\sum_{i=-1}^{N}A_i^{\nu
}(s),\label{eq2.12} \eeq with the definitions \beq Z^{\nu}(s) &=&
\sip \iinma\left\{\ln\left[z^{-\nu}I_{\nu} (z\nu
)\right]\right.\label{eq2.13}\\ & &\hspace{3cm}\left.
-\ln\left[\frac{z^{-\nu}}{\sqrt{2\pi\nu}}\frac{e^{\nu\eta}}
{(1+z^2)^{\frac 1 4}}\right]-\sum_{n=1}^N \frac{D_n (t)}{\nu
^n}\right\},\nn \eeq and \beq A_{-1}^{\nu } &=& \sip \iinma \ln
\left(z^{-\nu} e^{\nu\eta}\right),\label{eq2.14}\\ A_{0}^{\nu }
&=& \sip \iinma \ln (1+z^2) ^{-\frac 1 4},\label{eq2.15}\\
A_{i}^{\nu } &=& \sip \iinma \left(\frac{D_i (t)}{\nu
^i}\right).\label{eq2.16} \eeq

The essential idea is conveyed here by the fact that the
representation (\ref{eq2.12}) has the following anticipated
properties. First, by considering the asymptotics of the integrand
in eq.~(\ref{eq2.13}) for $z\to ma/\nu$ and $z\to\infty$, and by
considering the behaviour of $Z^\nu (s)$ for $\nu\to \infty$,
which is $\nu^{-2s-N-1}$, it can be seen that the function \beq
Z(s) =\sulnu (2l+1) Z^{\nu } (s)\label{eq2.17} \eeq is analytic on
the half plane $(1-N)/2<\Re s $. (The integral alone has poles at
$s=k\in \nats$ coming from the $z\to ma/\nu$ region which is seen
by writing $[(z\nu / a) ^2 -m^2]^{-s} = (-1)^j (\Gamma (1-s)/\G
(j+1-s)) (d^j/d(m^2)^j) [(z\nu / a) ^2 -m^2]^{-s+j}$. But these
are cancelled by the zeroes of the prefactor.) For this reason, it
gives no contribution to the residue of $\zeta (s)$ in that range.
Furthermore, for $s=-k$, $k\in\nats_0$, $k< (-1+N)/2$, we have
$Z(s) =0 $ and, thus, it also yields no contribution to the values
of the zeta function at these points. This result means that the
heat kernel coefficients, see eqs.~(\ref{eq1.8}) and
(\ref{eq1.9}), are just determined by the terms $A_i(s)$ with \beq
A_i  (s) =\sulnu (2l+1) A_i ^{\nu } (s).\label{eq2.18} \eeq

As they stand, the $\aid$ in eqs.~(\ref{eq2.14}), (\ref{eq2.15})
and (\ref{eq2.16}) are well defined on the strip $1/2<\Re s<1$ (at
least). We will now show that the analytic continuation in the
parameter $s$ to the whole of the complex plane, in terms of known
functions, can be performed. Keeping in mind that $D_i (t)$ is a
polynomial in $t$ (see the examples in eq. (\ref{eq2.11a})), all
the $\aid$ are in fact hypergeometric functions, which is seen by
means of the basic relation \cite{grad65b} \beq _2F_1 (a,b;c;z)
=\frac{\Gamma (c)}{\Gamma (b)\Gamma (c-b)}\int_0^1dt\,\, t^{b-1}
(1-t)^{c-b-1}(1-tz)^{-a}.\label{eq2.19} \eeq

Let us consider first in detail $\amed$, $\anud$, and the
corresponding $A_{-1}(s)$, $A_0(s)$. One finds immediately that
\beq \amed &=& \frac{\mzs}{2\sqrt{\pi}}am\frac{\g s-\frac 1 2
\right)}{\Gamma (s)}~_2F_1 \left(-\frac 1 2,s -\frac 1 2;\frac 1
2;-\numr\right)\nn\\ & &-\frac{\nu}2 \mzs ,\label{eq2.20}\\ \anud
&=& -\frac 1 4 \mzs ~_2F_1 \left(1,s;1;-\numr\right)\nn\\ & =&
-\frac 1 4 m^{-2s} \left[1+\numr\right]^{-s},\label{eq2.21} \eeq
where in the last equality we have used that $_2F_1 (b,s;b;x)
=(1-x)^{-s}$. These representations show the meromorphic structure
of $A_{-1}^\nu(s),A_0^\nu(s)$ for all values of $s$.

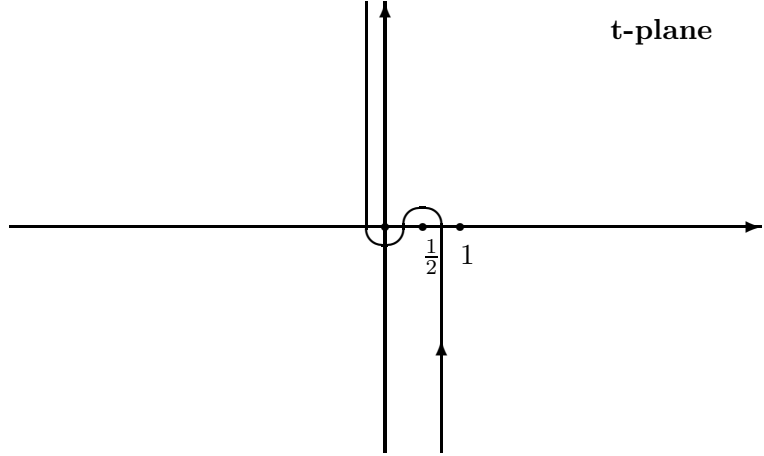
\begin{figure}[ht]
\setlength{\unitlength}{1cm}
\begin{center}
\begin{picture}(10,6.5)
\thicklines

\put(0,3){\vector(1,0){10}}
\put(5,0){\vector(0,1){6}}
\put(4.75,3){\line(0,1){3}}
\put(8.0,5.5){{\bf t-plane}}
\multiput(5.0,3)(.5,0){3}{\circle*{.1}}
\put(5.0,3){\oval(.5,.5)[b]}
\put(5.5,3){\oval(.5,.5)[t]}
\put(5.75,0){\line(0,1){3}}
\put(5.75,1){\vector(0,1){.5}}
\put(5.5,2.5){$\frac 1 2$}
\put(6.0,2.5){1}
\end{picture}

\caption{\label{fig2}Contour $\cac$ for eq.~(\ref{eq2.20}) }
\end{center}

\end{figure}

The next step is to consider the summation over $l$. For $\amed$
this is best done using a Mellin-Barnes type integral
representation of the hypergeometric functions, namely \beq _2F_1
(a,b;c;z) =\frac{\Gamma (c)}{\Gamma (a)\Gamma (b)}\frac 1 {2\pi i}
\int\limits_{\cac} dt\,\,\frac{\Gamma (a+t)\Gamma (b+t) \Gamma
(-t)}{\Gamma (c+t)}(-z)^{t},\label{eq2.22} \eeq where the contour
is such that the poles of $\Gamma (a+t) \Gamma (b+t) /\Gamma
(c+t)$ lie to the left of it and the poles of $\Gamma (-t)$ to the
right \cite{grad65b}. The contour involved in eq.~(\ref{eq2.20})
is shown in the Fig.~\ref{fig2}.

It is clear that one wishes to interchange the summation over $l$
and the integration in (\ref{eq2.22}) in order to arrive at a
Hurwitz zeta function, which is defined as
\beq \zeta_H (s;v)
=\sulnu (l+v)^{-s},  \ \ \ \  \Re s > 1.\label{eq2.23} \eeq

However, as is well known, one has to be very careful with this
kind of manipulations, what has been realized and explained with
great detail in \cite{weld86-270-79,eliz89-40-436,eliz95-28-617}.
Before interchanging the summation and integration one has to
ensure that the resulting sum will be absolutely convergent along
the contour $\cac$. Applying this criterium to $A_{-1}(s)$, \beq
A_{-1}(s) &=& \sulnu (2l+1)
\left[\frac{\mzs}{2\sqrt{\pi}}am\frac{\g s-\frac 1
2\right)}{\Gamma (s)} ~_2F_1 \left(-\frac 1 2,s-\frac 1 2;\frac 1
2;-\left(\frac{l+\frac 1 2}{ma}\right)^2\right)\right.\nn\\ &
&\hspace{4cm}\left.-\frac{l+\frac 1 2} 2 \mzs \right],\nn \eeq it
turns out that we may interchange the $\sum_l$ and the integral in
eq.~(\ref{eq2.22}) only if for the real part $\Re \cac$ of the
contour the condition $\Re \cac <-1$ is satisfied. However, the
integrand $\Gamma (-1/2+t) \Gamma (s-1/2+t)/\Gamma (1/2+t)$ has a
pole at $t=1/2$ (assume for the moment $\Re s \gg 1$ so that all
arguments go through). Thus the contour $\cac$ coming from
$-i\infty$ must cross the real axis to the right of $t=1/2$, and
then once more between $0$ and $1/2$ (in order that the pole $t=0$
of $\Gamma (-t)$ lies to the right of it), before going to
$+i\infty$ (see Fig.~2). That is, before interchanging the sum and
the integral we have to shift the contour $\cac$ over the pole at
$t=1/2$ to the left, which cancels the (potentially divergent)
second piece in $A_{-1} (s)$. This term, $-(l+1/2)m^{-2s}/2$, is
the result of the factor $\ln k^{-\nu}$ introduced in
eq.~(\ref{eq2.7}) to avoid contributions coming from the origin,
and its crucial importance is made explicit in the above step.

Closing then the contour to the left, we end up with the following
expression in terms of Hurwitz zeta functions (the contour at
infinity does not contribute, which is seen by considering the
asymptotics of the integrand) \beq A_{-1} (s)
&=&\frac{\rzs}{2\sqrt{\pi}\Gamma (s)}\sujnu \lll (ma)^{2j}\frac{\g
j+s-\frac 1 2\right)}{s+j} \zeta_H (2j+2s -2;1/2).\label{eq2.24}
\eeq

For $A_0(s)$, one only needs to use the binomial expansion in
order to find \beq A_0 (s) =-\frac{\rzs}{2\Gamma (s)}\sujnu \lll
(ma)^{2j}\Gamma (s+j) \zeta_H (2j+2s-1;1/2).\label{eq2.25} \eeq

The series are convergent for $|ma|<1/2$. Finally, we need to
obtain analytic expressions for the $A_i(s)$, $i\in\nats$. As is
easy to see, they are similar to the ones for $A_{-1} (s)$ and
$A_0(s)$ above. We need to recall only that $D_i (t)$,
eq.~(\ref{eq2.11}), is a polynomial in $t$, \beq D_i (t) =\suani
x_{i,b} t^{i+2b},\label{eq2.26} \eeq which coefficients $x_{i,a}$
are easily found by using eqs.~(\ref{eq2.10}) and (\ref{eq2.11})
directly. Thus the calculation of $\aid$ is essentially solved
through the identity \beq \iinma t^n &=& -\mzs
\frac{n}{2(ma)^n}\frac{\g s+\frac n 2 \right) \Gamma (1-s)}{\g
1+\frac n 2\right)}\nn\\ & & \times\nu ^n \left[1+\numr
\right]^{-s-\frac n 2}.\label{eq2.27} \eeq

The remaining sum may be done as mentioned for $A_0(s)$, and we
end up with \beq A_i (s) &=& -\frac{2\rzs}{\Gamma (s)}\sujnu \lll
(ma)^{2j} \zeta_H (-1+i+2j+2s;1/2) \nn\\ & &\hspace{1cm} \times
\suani x_{i,b} \frac{\g s+b+j+\frac i 2\right)}{\g b+\frac i
2\right)},\label{eq2.28} \eeq convergent once more for $|ma|<1/2$.
Restricting attention to the massless field, the asymptotic
contributions take the surprisingly simple form \beq A_{-1} (s)
&=& \frac{a^{2s}}{2 \sqrt{\pi} }
          \frac{\Gamma (s-1/2)}{\Gamma (s+1)} \zeta_H (2s-2;1/2),\nn\\
A_0 (s) &=& -\frac{a^{2s}}{2} \zeta_H (2s-1;1/2), \label{eq2.29}\\
A_i(s) &=& -\frac{2a^{2s}}{\Gamma (s)} \zeta_H (i-1+2s;1/2)
                \suani x_{i,b} \frac{\g s+b+j+\frac i
2\right)}{\g b+\frac i 2\right)}.\nn
\eeq

In summary, the analytical structure of the zeta function for the
problem considered is made completely explicit. Eq. (\ref{eq2.29})
allows for the direct determination of residues, function values
and derivatives at whatever values of $s$ needed. Clearly, all
ingredients can be easily dealt with by Mathematica and
calculations can be easily automized. The remaining piece to
obtain the full zeta function, eq. (\ref{eq2.17}) and
(\ref{eq2.13}), is suitable for numerical evaluation for values of
$s$ in the half plane mentioned. Furthermore, an analytical
treatment of $Z'(0)$ is possible, which leads to the determinant
$\zeta ' (0)$ of the Laplacian on the ball in terms of elementary
Hurwitz \zfs \cite{bord96-179-215,bord96-182-371}. For various
different approaches in a similar context see also
\cite{dowk96-13-1,dowk96-366-89,barv92-219-201,falo96-37-5805,falo98-39-532},
with possible applications in quantum cosmology
\cite{espo94b,espo97b}.

\section{Calculation of heat-kernel coefficients via special cases}
\label{sec4}
\setcounter{equation}{0}
Let us now come to the first application of our analysis in Section
\ref{sec3.1},
namely the calculation of heat-kernel coefficients for Laplace-like operators
on smooth manifolds with smooth boundaries.
Our main
emphasis is on the determination of the boundary contribution to the
heat kernel coefficients.
This is motivated because it is here that the special case of
the ball gives rich information. But it is also justified
because the calculation of the volume part is nowadays nearly automatic
\cite{avra91-355-712,full88-310-583,ven98-15-2311} and these terms do
not depend on the \bcs \cite{seel69-91-889}
and are thus known already for all problems
to be dealt with.

Concerning the relevance of heat-kernel coefficients, in the
Introduction we have already mentioned that a knowledge of these
coefficients is equivalent to a knowledge of the one-loop
renormalization equations in various theories
\cite{wies94-233-125}, which provides one reason for the
consideration of heat-kernel coefficients in physics. In addition,
if an exact evaluation of relevant quantities is not possible,
asymptotic expansions (with respect to inverse masses, slowly
varying background fields, high temperature,...) are often very
useful and naturally given in terms of heat-kernel coefficients.
An especially important link between physics and mathematics is
provided by index theorems \cite{gilk95b}, again, with the
well-known connection to the heat equation. But, in mathematics,
the interest extends to basically all of Geometric Analysis,
including analytic torsion \cite{fran35-173-245,ray71-7-145},
characteristic classes \cite{gilk95b}, sharp inequalities of
borderline Sobolev and Moser-Trudinger type
\cite{bran95-347-3671}, etc.

We will start the Section by giving what can be named the general
form of the \hks for Dirichlet and Robin boundary conditions. (We
will explain the approach for these boundary conditions; comments
on various other boundary conditions are given at the end of the
Section.) As we will see these are built from certain geometrical
invariants with unknown numerical coefficients. Relations between
the unknown coefficients can be derived by conformal
transformation techniques most systematically used by Branson and
Gilkey \cite{bran90-15-245}. However, in order to determine the
numerical coefficients, additional information is needed. The
product formula gives a certain subset of the numerical
coefficients but in general by far not enough to complete the
calculation by using the conformal techniques
\cite{bran90-15-245}. Especially the group of terms containing the
extrinsic curvature is not even touched by the product formula and
our calculation on the ball will turn out to be very valuable.
Combined with the conformal techniques, the application of index
theorems, and additional examples, the conglomerate of all methods
allows for the determination of (at least) the leading \hks for
all classical boundary conditions.

Let us clearly state that the coefficients we are going to "determine",
namely the coefficients up to $a_{3/2}$ for Dirichlet and Robin boundary conditions,
are known already for a long time, see. f.e.~\cite{bran90-15-245}. It is just
to explain the ideas in detail that we have chosen the first coefficients
and the simplest boundary conditions.
But in fact, the method has been successfully applied to $a_{5/2}$
containing more than $100$ terms \cite{kirs98-15-5}
(see also \cite{bran97-11-39}), showing the effectiveness of the
procedure employed.

\subsection{General form of the coefficients
for Dirichlet and Robin boundary conditions}
\label{sec4.1}
In order to be as self-contained as possible, in the following
we are going to summarize some basic properties of heat-kernel coefficients
on manifolds with boundary. We follow \cite{bran90-15-245}.

Let $\cam$ be a compact $D=(d+1)$-dimensional Riemannian manifold with
boundary $\pa \cam$. Let $V$ be a smooth vector bundle over $\cam$
equipped with a connection $\nabla^V$ and finally let $E$ be an
endomorphism of $V$. Our interest is then in Laplace-type
operators of the form (\ref{eq1.1g}),
\beq P= -g^{ij} \nabla_i^V
\nabla_j^V - E . \label{eq3.1} \eeq

Let us mention that every second order elliptic differential
operator on $\cam$ with leading symbol given by the metric can be
put in this form. We will see this explicitly below, starting with
eq. (\ref{eq3.16}).

If there is no boundary, it is well known that $P$ defines a
symmetric operator. If, however, there is a boundary present,
Green's theorem says \beq (v,Pw)_{L^2} - (Pv,w)_{L^2} = \int _\cam
dx
  (v ^\dagger Pw -(Pv)^\dagger w ) = \int _{\pa \cam} dy
     (  v_{;m}^\dagger w -v^\dagger  w_{;m}) ,\label{eq3.1a}
\eeq with $dx$ and $dy$ the volume elements on $\cam$ and
$\pa\cam$, and $v_{;m}$ the normal covariant derivative of $v$
with respect to the {\it exterior} normal $N$ to the boundary
$\pa\cam$. In order that $P$ be symmetric one has to impose
boundary conditions. Obvious possibilities are the classical
Dirichlet or Robin boundary conditions, \beq \cab ^- \phi \equiv
\phi|_{\partial \cam } \quad \mbox{and} \quad \cab_S^+ \phi \equiv
\left( \phi_{;m} -S\phi \right) |_{\partial \cam}   ,
\label{eq3.2} \eeq with $S$ a hermitian endomorphism of $V$
defined on $\pa\cam$. Clearly, in these cases the argument of the
integral over the boundary in (\ref{eq3.1a}) vanishes and, with
these boundary conditions, a symmetric operator $P$ is defined.

Another possibility is to impose a mixture of Dirichlet and Robin
boundary conditions. In order to do this, a suitable splitting of
$V= V_- \oplus V_+$ is needed and in $V_-$ one imposes Dirichlet
and in $V_+$ Robin boundary conditions. Also here, the integrand
itself vanishes.

However, a further possibility is that the integrand in
(\ref{eq3.1a}) does not vanish but equals a boundary divergence.
This condition involves tangential derivatives and, in the
mathematical literature, it is sometimes referred to as oblique.
Although these last \bcs have been the subject of classical
analysis (see, f.e. \cite{trev80b,egor91b,kran92b}), very little
is known about the associated heat equation asymptotics. We will
comment on the application of the techniques to all these \bcs at
the end of the section, but concentrate now on conditions
(\ref{eq3.2}).

To have a uniform notation, we set $S=0$ for Dirichlet boundary
conditions and write $B_S^\mp$. If $F$ is a smooth function on
$\cam$, there is an asymptotic series as $t \to 0$ of the form
\beq \mbox{Tr} _{L^2} \left( F e^{-tP} \right) \approx \sum_{n\geq
0 } t^{n-\frac D 2 } a_n (F,P, \bou ) , \label{eq3.3} \eeq where
the $a_n (F,P, \bou)$ are locally computable \cite{gilk95b}.

The use of the smearing function $F$ is important for at least
three reasons. First, near the boundary the heat-kernel behaves
like a distribution and, by studying $\mbox{Tr} _{L^2} ( F
e^{-tP})$, this local behaviour is recovered. As an example
consider $\pa\cam = \emptyset$. For $F=1$ volume divergences are
integrated away, which is not possible if a smearing function is
present. A second reason is that for the functorial formalism to
be described \cite{bran90-15-245}, the smearing function is at the
heart of this method. Finally, it is exactly this {\it smeared}
coefficient appearing in the integration of conformal anomalies
that is relevant for several physical applications
\cite{blau89-4-1467,dowk86-33-3150,dowk88-38-3327,dowk89-327-267,dowk90-31-808,blau88-209-209,buch86-44-828,gusy87-46-1832,wipf95-443-201,gamb84-157-360}.

To state the general form of the coefficients in (\ref{eq3.3}),
let us introduce some notation. Here and in the following
$F[\cam]=\mbox{Tr}\,\,\int_{\cam}dx\, F(x) $ and $F[\partial \cam]
= \mbox{Tr}\,\,\int_{\partial \cam}\, dy F(y) $, with $\mbox{Tr}$
the fiber trace. In addition, "$;$" denotes differentiation with
respect to the Levi-Civita connection of $\cam$ and "$:$"
covariant differentiation tangentially with respect to the
Levi-Civita connection of the boundary. Furthermore, $\Om$ is the
curvature of the connection $\nabla^V$, $[\nabla_i^V , \nabla_j^V]
= \Om_{ij}$, and $R_{ijkl}$, $R_{ij}$, $R$, are as usual Riemann
tensor, Ricci tensor and Riemann scalar. Finally let $N^{\nu} (F)
= F_{;m...}$ be the $\nu^{th}$ normal covariant derivative. Then
there exist local formulae $a_n (x,P)$ and $a_{n,\nu} (y,P, \bou)$
so that \cite{gilk83-36-85} \beq a_n (F,P,\cab _S^\mp ) = \{Fa_n
(x,P)\} [\cam] +
 \{  \sum_{\nu = 0}^{2n-1} N^\nu (F) a_{n,\nu} (y,P,\cab_S^\mp )\}
   [\partial \cam].
\label{eq3.4}
\eeq

Important homogeneity properties follow from the Seeley calculus
\cite{seel69-91-889}, so one has for $0<c\in\reals$
\cite{gilk83-36-85} \beq a_n (x,c^{-2} P  ) =c^{-2n} a_n (x,P ),
\quad a_{n,\nu } (y,c^{-2} P , c^{-1} \bou ) = c^{-(2n-\nu )}
a_{n,\nu}
      (y, P , \bou ) .\label{eq3.4a}
\eeq

Physicists would say that $P$ carries dimension {\it
length$^{-2}$}, thus in order that $e^{-tP}$ makes sense, $t$ has
to have dimension of {\it length$^2$}. Then, eq.~(\ref{eq3.3}) is
dimensionless and $a_n (F,P, \bou )$ must have dimension of {\it
length$^{-D+2n}$}. As a result, $a_n (x,P_\cab)$ has dimension of
length to the power $2n$ and $a_{n,\nu} (y,P_\cab , \bou )$ to the
power $2n-\nu$ which is equivalent to the above.

The interior invariants $a_n (x,P)$ are built universally and polynomially from
the metric tensor, its inverse, and the covariant derivatives of $R,\Om$, and
$E$. By Weyl's work on the invariants of the orthogonal group
\cite{weyl15-39-1}, these polynomials can be formed using only tensor
products and contraction of tensor arguments (indices).
If $A$ is a monomial term of $a_n (x,P)$ of degree $(k_R,k_\Om ,k_E)$ in
$(R,\Om ,E)$, and if $k_\nabla$ explicit covariant derivatives appear in
$A$, then by the homogeneity property of $a_n (x,P)$,
\beq
2 (k_R + k_\Om +k_E ) + k_\nabla = 2n .\nn
\eeq

When considering the boundary invariants $a_{n,\nu} (y,P , \bou)$
we must also introduce the second fundamental form $K_{ab} =
(\nabla_{e_a} e_b, N)$, $K=K_a^a$, where $\{e_1,...,e_d\}$ is an
orthonormal frame of $T(\pa\cam)$, the tangent bundle of the
boundary and, when considering Robin boundary conditions, we must
also consider the tensor $S$. Given that these are defined only at
the boundary, we only differentiate $\{K,S\}$ tangentially. We use
Weyl's \cite{weyl15-39-1} theorem again to construct invariants.
The structure group now is $O(D-1)$, and the normal $N$ plays a
distinguished role. If $A$ is a monomial term of $a_{n,\nu} (y,P,
\bou)$ of degree $(k_R,k_\Om ,k_E,k_K,k_S)$ in $(R,\Om ,E,K,S)$,
and if $k_\nabla$ explicit covariant derivatives appear in $A$,
then once more by homogeneity \beq 2 (k_R + k_\Om +k_E ) +k_K
+k_S+ k_\nabla = 2n -\nu .\nn \eeq

By constructing a basis for the space of invariants of a given
homogeneity, we write down the following general form of the \hks
\cite{bran90-15-245}, \beq a_0 (F,P, \bou) &=& (4\pi
)^{-D/2} F[\cam ],\label{eq3.5}\\ a _{1/2} (F,P, \bou)  &=& \delta
(4\pi )^{-d/2} F[\partial \cam],\label{eq3.6}\\ a _1( F,P, \bou)
&=& (4\pi )^{-D/2} 6^{-1} \left\{
 ( 6 FE + FR ) [\cam ] + (b_0 FK +b_1 F_{;m} + b_2 F S ) [\partial \cam]
\right\},\label{eq3.7}\\
a_{3/2} (F,P, \bou) & =& {\delta\over96 (4\pi )^{d/2}} \{\!
     F\big(c_0 E +c_1 R\! +c_2 R_{mm} +c_3 K^2\! +c_4 K_{ab} K^{ab}
      +  c_7 SK\! +c_8 S^2\!\big)\nn\\
& &\phantom{\delta (4\pi )^{-d/2} 96^{-1}}
            +F_{;m} (c_5 K +c_9 S) +c_6 F_{;mm} \}[\partial \cam] .
\label{eq3.8} \eeq

All numerical constants involved have the very important property
of being independent of the dimension $D$ \cite{gilk83-36-85}. A
direct proof has been given in \cite{bran90-15-245} and this goes
as follows. First one shows a product formula for heat-kernel
coefficients. Let $\cam = \cam _1 \times \cam _2$ and $P=P_1
\otimes 1 + 1\otimes P_2$ and $\partial \cam _2  = \emptyset $ and
let $S$ only depend on coordinates in $\cam_1$. Then \beq a_{n}
(x,P ) &=& \sum_{p+q=n} a_{p} (x_1,P_1 ) a_q (x_2,P_2),\nn\\
a_{n,\nu} (y,P,\cab _S ^\mp ) &=& \sum_{p+q=n} a_{p,\nu}
(y_1,P_1,\cab _S^\mp ) a_q (x_2,P_2).\label{eq3.11} \eeq

This is a purely formal computation because by separation of
variables the heat-kernel of the operator $P$ gets the product of
the heat-kernels of $P_1$ and $P_2$ and the result follows by just
comparing powers of $t$.

To avoid the appearance of factors of $\sqrt{4\pi}$ normalize for the
moment $a_0 (x,P) =1$.
Application of formula (\ref{eq3.11}) to
$(\cam_2 , P_2)= (S^1,-\pa^2/\pa \theta^2)$
leads to $a_n (x_1,P_1) = a_n ((x_1,\theta),$ $P)$ and
$a_{n,\nu} (y_1,P_1, \bou)= a_{n ,\nu } ((y_1,\theta),P,\bou)$,
because
$a_0 (\theta ,P_2) =1$ and $a_q (\theta , P_2 ) =0$ for $q>0$.
However,
invariants formed by contractions of indices are restricted from
$\cam_1 \times S^1$ to $\cam_1$ by restricting the range of summation,
but have the same appearance. This shows that the numerical constants
are
independent of the dimension.

We are thus left with the task of the determination of the
universal numerical constants by whatever method. An obvious
(rich) source of information are special case calculations. Let us
discuss this in detail for the example of the ball in order to
motivate the calculation of the coefficients for this setting. The
way to deal with the three dimensional ball and Dirichlet \bcs has
been shown in Section \ref{sec3.1}. In Section \ref{sec3.2} we
will show how these results can be generalized to arbitrary
dimensions $D=d+1$. In addition, the way to deal with Robin \bcs
with a constant endomorphism $S$ is explained. For these manifolds
we will have $K_a^b=\de _a^b$. As a result we get $K=d$,
$K_{ab}K^{ab} = d$, $K^2=d^2$, and so on. The polynomials, traces
and contractions of $K_{ab}$ give a polynomial in the dimension
$d$. In addition, for the example of the ball, we have
$R_{abce}=0$, and we have chosen $P=-\De_\cam$, thus $E=0$.
Finally, we included no smearing function and have $F=1$. In this
setting, restricting (\ref{eq3.6}) to the ball, we have \beq
a_{1/2} (1,-\De_\cam, \bou ) = \de (4\pi) ^{-d/2} |S^d|,\nn \eeq
with the volume $|S^d| = 2\pi^{(d+1)/2}/\G ((d+1)/2)$ of the
$d$-sphere. Calculating the heat-kernel coefficients explicitly
one will find the "unknown" numerical constants $\de$ and then
knows $a_{1/2} (F,P,\bou)$ for an arbitrary manifold by eq.
(\ref{eq3.6}). Passing on to $a_1$, on the ball we have \beq a_1
(1, -\De_\cam, \bou ) = (4\pi) ^{-D/2} 6^{-1} |S^d|  \left( b_0 d
+b_2 S \right) .\nn \eeq

Just by comparing powers of $d$ and $S$ one can determine $b_0$
and $b_2$ from the explicit $a_1 (1, -\De_\cam $ $\bou )$ on the
ball. Let us stress that, if we include a smearing function $F(r)$
into the formalism, then $F_{;m}= (d/dr) F(r)$ is the derivative
with respect to the exterior derivative, and we would have
determined also $b_1$. By application of (\ref{eq3.7}) we see that
one can get $a_1 (F,P,\bou )$ by just having the result on the
ball.

Continuing with the same argumentation, in eq. (\ref{eq3.8}) one
can determine $c_3,c_4,c_7$ and $c_8$ for $F=1$ and furthermore
$c_5,c_9$ and $c_6$ including an $F(r)$. Let us stress that $c_3$
and $c_4$ both can be determined only because we will perform our
calculation in arbitrary dimension (this observation gets more
important for the higher coefficients). Thus only $3$ of $10$
unknowns are left and it gets clear that the special case
calculation chosen contains rich information. However, one also
realizes, and this was, of course, clear from the beginning that
the example can not determine the full coefficients, because $E$
and the Riemann tensor vanishes. Both aspects can be improved a
bit by including a mass (as we have done) and by dealing with a
so-called bounded generalized cone \cite{bord96-182-371}. But the
information obtained thereby is also very easily obtained by an
application of the product formula (\ref{eq3.11}) and we will not
give details of this generalization here. However, product
manifolds have vanishing normal components of the Riemann tensor
(their appearance starts with $a_{3/2}$) and the corresponding
universal constants have to be determined by different means.

An extremely effective method to continue at this point is to study
the functorial properties of the \hks \cite{bran90-109-437,bran90-15-245}.
We summarize here the main points in order to allow for a self-consistent
reading. Consider the one-parameter family of differential operators
\beq
P(\ep ) = e^{-2\ep F} P \label{eq3.12}
\eeq
and boundary operators
\beq
\bou (\ep ) = e^{-\ep F} \bou.\label{eq3.13}
\eeq

Here, $\ep$ is a real-valued parameter and $F$, as before, is a
function. Eq. (\ref{eq3.13}) guarantees that the boundary
condition remains invariant along the one-parameter family of
operators (see below, eq. (\ref{eq3.18a})). Then one might ask
what is the dependence of the \hks on the parameter $\ep $. The
relevant transformation behaviour is described by the following \\
\\
{\bf Lemma:}
\beq
(a) & &\frac d {d\ep}\left.\right|_{\ep =0} a_n (1,P(\ep ) ,\bou (\ep ) )
                    = (D-2n)
                     a_n (F,P, \bou), \label{eq3.14}\\
(b) & &\mbox{If }D=2n+2,\mbox{ then }
 \frac d {d\ep}\left.\right|_{\ep =0} a_n \left(e^{-2\ep f} F,P(\ep ),
                           \bou (\ep )
                           \right)=0.
                         \label{eq3.15}
\eeq
\ph{$$}\\

Proceeding formally, (a) is proven by considering \beq & &\frac d
{d\ep}\left.\right|_{\ep =0} \mbox{Tr}_{L^2} \left( e^{-tP(\ep
)}\right) =-t \mbox{Tr}_{L^2}\left(\left[ \frac d
{d\ep}\left.\right|_{\ep =0}P(\ep )\right]e^{-tP}\right) \nn\\
&=&2t\mbox{Tr}_{L^2} \left(FPe^{-tP} \right) = -2t \frac{\pa}{\pa
t} \mbox{Tr}_{L^2} \left(F e^{-tP}\right),\nn \eeq and comparing
powers of $t$ in the asymptotic expansion. For the necessary
justification of the analytic steps see \cite{gilk83-36-85}. Part
(b) is much the same by starting with $(d/d\ep)|_{\ep =0}
\mbox{Tr}_{L^2} (e^{-2\ep f}F e^{-tP(\ep )})$.

This Lemma will be applied in Section \ref{sec4.1b}; let us here
only explain the way it determines universal constants. First, we
obviously need the \hks for the operator $P(\ep )$, so let us see
how the family (\ref{eq3.12}) of operators can be generated. Let
$P$ be an arbitrary second order differential operator with
leading symbol given by the metric tensor. In local coordinates we
have \beq P=- \left(  g^{ij} \frac{\pa^2}{\pa x^i\pa x^j}+
           P^k \frac{\pa}{\pa x^k}  +Q  \right). \label{eq3.16}
\eeq
Obviously, $P(\ep )$ is obtained by defining
\beq
g^{ij} (\ep ) = e^{-2\ep F}g^{ij} , \quad
P^k (\ep ) = e^{-2\ep F} P^k, \quad Q(\ep ) = e^{-2\ep F} Q.\nn
\eeq
In order to obtain the \hks of $P(\ep )$ in the form
(\ref{eq3.5})---(\ref{eq3.8}), write $P$ invariantly in the form
(\ref{eq3.1}), $P=-(g^{ij}\nabla_i^V \nabla_j^V +E)$. Let $\om_l$ be the
connection $1$ form and consequently
$\Om_{ij} = [\nabla_i^V,\nabla_j^V]= \om_{j,i} -\om_{i,j}
+\om_i \om_j -\om_j \om_i $
with "," the partial derivative.
Comparing the two different representations of
$P$ one finds
\beq
\om_l &=& \frac 1 2 g_{il} \left( P^i +g^{jk} \G^i_{jk}\right),\label{eq3.17}\\
E&=& Q-g^{ij} \left( \om_{i,j} +\om_i\om_j-\om_k \G^k_{ij}\right),
\label{eq3.18}
\eeq
with the Christoffel symbols $\G^i_{jk}$.
As a result, this defines
\beq
\om_l (\ep) &=& \om_l +\frac 1 2 \ep (2-D) F_{;l} \label{eq3.18a}\\
\Om_{ij} (\ep ) &=& \Om_{ij}\label{eq3.18b}\\
E(\ep ) &=& e^{-2\ep F}(
       E+\frac 1 2 (D-2) \ep \De_\cam F +\frac 1 4 (D-2)^2 \ep^2 F_{;k}F_;^k).
\label{eq3.18c} \eeq

This shows that the leading $a_n(f,P(\ep ),\bou (\ep ))$ are given
by eqs. (\ref{eq3.5})---(\ref{eq3.8}) once the above definitions
are used and once all geometrical tensors and covariant
derivatives are calculated with respect to the metric $g_{ij} (\ep
)$ (for many useful relations see \cite{bran90-15-245}). Let us
here only mention that Dirichlet \bcs are obviously conformally
invariant, and that with \beq S(\ep ) = e^{-\ep F} \left( S - \ep
\frac{D-2} 2 F_{;m}\right)\nn \eeq the same holds for Robin
conditions. This is seen immediately from eq.~(\ref{eq3.18a}) and
the boundary condition (\ref{eq3.2}). As a result, $(d/d\ep
)|_{\ep =0} a_k (1,P(\ep ), \bou (\ep )) $ will have the same
appearance as $a_k (F,P, \bou)$, and Lemma (\ref{eq3.14}) will
give relations among the universal constants, as well as
(\ref{eq3.15}) does. To make it very clear, look at $a_{3/2}
(F,P,\bou )$. Then $(d/d\ep) E(\ep)$ contains a term $F_{;mm}$
being part of $\Delta F$, see (\ref{eq3.18c}). Eq. (\ref{eq3.14})
then states that the numerical constant $c_0$ is connected with
$c_6$ (more invariants are involved however). Let us stress
already here, that due care must be taken that only {\it
independent} terms are compared in eqs.~(\ref{eq3.14}) and
(\ref{eq3.15}) and that partial integrations (or more involved
manipuliations) may be necessary to see that apparently
independent terms are actually dependent.

On its own, the functorial method is unable to determine the
coefficients fully. But, given a subset of numerical coefficients,
found f.e. by special case calculations, the method provides the
required information with relative ease. For that reason, we start
by applying the product Lemma (\ref{eq3.11}) and, by calculating
the \hks for the Laplacian on the ball, determine (a subset of
the) universal constants and complete the calculation by use of
the functorial properties.

In the last step it will turn out that the generalization of the
calculations in Section \ref{sec3.2} to the smeared zeta function
is essential. This is, as seen in eq.~(\ref{eq3.14}), because the
functorial techniques (apart from other things) yield relations
between the smeared and non-smeared case; the information one can
get on the "smeared side" is crucial to find the full
"non-smeared" side. The way this can actually be done is explained
in \cite{dowk9803094}.

\subsection{Scalar field on the $D$-dimensional ball}
\label{sec3.2}

As briefly explained, for the applications of our calculations to
the heat-equation asymptotics, it will be very important that
results in arbitrary dimensions are available. For that reason,
let us briefly explain how the three dimensional calculation of
Section \ref{sec3.1} is generalized to arbitrary dimension $D=d+1$
\cite{bord96-182-371}. We put the radius of the ball $a=1$, the
dependence of the results on the radius are easily recovered by
dimensional arguments.

Instead of (\ref{eq2.3}),
the nonzero eigenmodes of $\De$ that are finite at the origin have
eigenvalues $-\al^2$ and are of the form
\beq
{J_{\nu}(\al r)\over r^{(d-1)/2}}\,Y_{l+D/2}(\Om), \label{eq2.34}
\eeq
where the spherical harmonics satisfy
\beq
\De_{\can} Y_{l+D/2}(\Om)=-\la_l^2 Y_{l+D/2}(\Om)\label{eq2.35}
\eeq
with $\De_\can$ the Laplacian on the sphere, and
\beq
\nu^2= \la_l^2+(d-1)^2/4  . \label{eq2.36}
\eeq

The eigenvalues are \cite{erde55b} \beq \lambda_l^2 = l (l+d-1) =
\left( l+\frac{d-1} 2 \right)^2
             -\left(\frac{d-1} 2 \right) ^2 , \quad l\in\nats_0,\nn
\eeq
such that
\beq
\nu^2 = \left( l+\frac{d-1} 2 \right)^2 ,\nn
\eeq
and they have degeneracy
\beq
d(l) = (2l+d-1) \frac{(l+d-2)!}{l!(d-1)!}.\nn
\eeq

Let us next impose the boundary conditions and this time let us
consider Dirichlet as well as generalized Neumann (or Robin)
boundary conditions. In the notation of, for example,
\cite{levi98-8-35}, these read explicitly \beq J_{\nu}(\al) =0
\label{eq2.39} \eeq for Dirichlet and \beq u J_{\nu} (\al ) +\al
J' _{\nu} (\al ) =0 \label{eq2.40} \eeq for Robin, with $u\in
\reals$. In the following we will write $u=1-\frac D 2 -\beta$,
such that $\beta =0$ corresponds to Neumann boundary conditions
and $\beta$ is to be identified with the endomorphism $S$.

As we have seen in the three dimensional calculation,
the angular momentum sum leads to a zeta function of the type,
\beq
\zb (s) = \sum d(\nu) \nu^{-2s}, \label{eq2.41}
\eeq
and we anticipate this to be the central object to state the asymptotic
contributions. Its definition is clearly motivated by the calculation in
$3$ dimensions where $2\zh (2s-1;1/2)$ plays the role of $\zb (s)$.

Our first aim will be to express the whole zeta function on the ball
\beq
\ze (s) = \sum \alpha ^{-2s},\nn
\eeq
as far as possible in terms of this quantity. That is, we seek to replace
analysis on the ball by that on the sphere.

Following the analysis of the previous section for Dirichlet
boundary conditions, 
the starting point is the representation of the zeta function in
terms of a contour integral \beq \zeta (s)= \sum \ent \ikma\ln
J_{\nu} (k),\label{eq2.42} \eeq where the anticlockwise contour
$\gamma$ must enclose all the solutions of (\ref{eq2.39}) on the
positive real axis.

As we have already seen, it is very useful to split the zeta
function into two parts,
\beq \zeta (s) =Z(s)
+\sum_{i=-1}^{N}A_i(s).\label{eq2.43} \eeq

Performing identical steps as before, the different pieces are
determined to be
\beq Z(s) &=& \sip \sum d(\nu )
\int\limits_0^{\infty} dz\,\,(z\nu )^{-2s}\frac{\pa}{\pa z}
\bigg(\ln\left(z^{-\nu}I_{\nu} (z\nu )\right)\label{eq2.44}\\ &
&\hspace{3cm}
-\ln\bigg[\frac{z^{-\nu}}{\sqrt{2\pi\nu}}\frac{e^{\nu\eta}}
{(1+z^2)^{\frac 1 4}}\bigg]-\sum_{n=1}^N \frac{D_n (t)}{\nu
^n}\bigg),\nn \eeq and \beq A_{-1} (s) &=& \frac 1 {4\sqrt{\pi}}
\frac{\Gamma \left(s-\frac 1 2\right)} {\Gamma (s+1)} \zb \left(
s-1/2 \right),\label{eq2.45}\\ A_0 (s) &=& -\frac 1 4 \zb (s),
\label{eq2.46}\\ A_i (s) &=& -\frac 1 {\Gamma (s)} \zb \left(
s+i/2 \right) \sum_{b=0}^i x_{i,b} \frac{\g s+b+i/2\right)} {\g
b+i/2\right)}.\label{eq2.47} \eeq

These results can be read off from eqs. (\ref{eq2.13}) and
(\ref{eq2.17}) once the definition (\ref{eq2.36}) for $\nu^2$ is
used and $2\zh (2s-1;1/2)$ is replaced by $\zeta _{\can} (s)$,
eq.~(\ref{eq2.41}). The
function $Z(s)$ is analytic on the strip $(d-1-N)/2 < \Re s$,
which may be seen by considering the asymptotics of the integrand
in eq.~(\ref{eq2.44}).

As is clearly apparent in eq.~(\ref{eq2.45})--(\ref{eq2.47}),
sphere contributions are separated from radial ones. Again, the
analytical structure is made very explicit and the result is very well
organized by the introduction of the zeta function $\zeta _\can (s)$.

In order to treat Robin boundary conditions, only a few changes are necessary.
In addition to expansion (\ref{eq2.9}) we need
\cite{olve54-247-328,abra70b}
\beq
I_{\nu} ' (\nu z )\sim
\frac 1 {\sqrt{2\pi \nu}} \frac{e^{\nu \eta} (1+z^2)^{1/4}} z
\left[1+\sum_{k=1}^{\infty}\frac{v_k(t)}{\nu^k}\right], \label{eq2.48}
\eeq
with the $v_k (t)$ determined by
\beq
v_k (t) = u_k (t) +t (t^2 -1) \left[
\frac 1 2 u_{k-1} (t) +t u_{k-1} ' (t) \right].\label{eq2.49}
\eeq

The relevant polynomials analogous to the $D_n (t)$,
eq.~(\ref{eq2.11}), are defined by \beq \ln\left[
1+\sum_{k=1}^{\infty}\frac{v_k (t)}{\nu^k} + \frac{ 1-D/2 -\beta
}{\nu} t \left(1+\sum_{k=1}^{\infty} \frac{u_k (t)} {\nu^k}\right)
\right] \sim  \sum_{n=1}^{\infty} \frac{M_n
(t)}{\nu^n}\label{eq2.50} \eeq and have the same structure, \beq
M_n (t) =\sum_{b=0}^n z_{n,b}\, t^{n+2b}.\label{eq2.51} \eeq

One may again introduce a split as in eq.~(\ref{eq2.43})
with $A_{-1}^R (s) =A_{-1} (s)$ and  $A_0^R(s) =-A_0 (s)$, where
the upper index $R$ indicates that these are the results for Robin
boundary conditions.
The $A_i^R (s)$ are given by eq.~(\ref{eq2.47}) once the
$x_{i,a}$ is replaced by $z_{i,a}$.
In addition one finds
\beq
Z^{R} (s) &=& \sip \sum d(\nu )\int\limits_0^{\infty}
dz\,\,(z\nu )^{-2s}\frac{\pa}{\pa z}
\bigg(\ln\left( (1-D/2-\beta ) I_{\nu} (z\nu
) + z\nu I' _\nu (z \nu  ) \right)\nn\\
& &\hspace{3cm}
-\ln\bigg[\sqrt{\frac{\nu}{2\pi}}e^{\nu\eta}
(1+z^2)^{\frac
1 4}\bigg]-\sum_{n=1}^N \frac{M_n (t)}{\nu ^n}\bigg).\label{eq2.52}
\eeq

The remaining task is the analysis of the meromorphic structure of
the zeta function $\zeta_\can (s)$. It reads explicitly \beq \zb
(s) = \slnu (2l+d-1) \frac{(l+d-2)!}{l!(d-1)!} \left(l+
                  \frac{d-1} 2 \right) ^{-2s}.\label{eq2.53}
\eeq

Writing \beq d(l) = \left(    \begin{array}{c}
               l+d-1 \\
                 d-1
               \end{array} \right)
       + \left(    \begin{array}{c}
             l+d-2 \\
             d-1
               \end{array} \right) ,  \nn
\eeq
it is seen immediately that this is a sum of Barnes zeta functions
\cite{barn03-19-374,barn03-19-426,dowk94-35-4989}
defined as
\beq
\zba (s,a) \sum_{\vec m =0}^\infty \frac 1 {(a+m_1 + ... + m_d ) ^s}
                = \slnu
        \left(    \begin{array}{c}
               l+d-1 \\
                 d-1
               \end{array} \right)
          (l+a)^{-s} .  \label{eq2.54}
\eeq

In detail, one finds \beq \zb (s) = \zba \left( 2s, \frac{d+1} 2
\right) +\zba \left( 2s, \frac{d- 1} 2 \right). \label{eq2.55}
\eeq

Its residues are determined in the following way. Using the
integral representation \beq \zeb (s,c) =\frac {i\Gamma
(1-s)}{2\pi} \int_L dz\,\,\frac{e^{z\left( d/ 2 -c\right)}
(-z)^{s-1}} {2^d\sinh^d\big(z/2\big)},\label{eq3.20} \eeq where
$L$ is the Hankel contour, one immediately finds for the base
function \beq \zb (s) &=&   \frac{i\Gamma (1-2s)}{2\pi} 2^{2s+1-d}
\int_L dz  \,\, (-z)^{2s-1}\frac{\cosh z}{\sinh ^d
z}\label{eq3.21}\\ &
=&\frac{i\Gamma(2-2s)}{2\pi(d-1)}2^{2s+1-d}\int_L dz
\,\,{(-z)}^{2s-2}{1\over\sinh^{d-1}z}.\nn \eeq

For the residues this yields ($m=1,2,...,d$) \beq \Res\zb (m/2) =
\frac{2^{m-d} \pold _{d-m}}{(d-1)(m-2)! (d-m)!} , \label{eq3.22}
\eeq with the $\pold _{\nu}$ defined through ({\it cf}
\cite{chod84-156-412}) \beq \left(\frac z {\sinh z}\right)^{d-1}
=\sum_{\nu =0}^{\infty} \pold _{\nu} \frac{z^{\nu}}{\nu
!}.\label{eq3.23} \eeq

Obviously $\pold _{\nu}=0$ for $\nu$ odd, so there are actually
poles only for $m=1,2,...,d$ with $d-m$ even. The advantage of
this approach is that known recursion formulas allow for an
efficient evaluation of the $D^{(n)}_\nu$ as polynomials in $d$,
\cite{norl22-43-121}.

Using eq.~(\ref{eq3.22}) in eqs.~(\ref{eq2.45})---(\ref{eq2.47})
we find, by the connection (\ref{eq1.8}) for the heat-kernel
coefficients $\hem_ {k/2}$ with Dirichlet boundary conditions,
\beq \frac{(4\pi)^{D/2}}{|S^d|} \hem _{k/2} &=&
\frac{(d-k-1)}{(d-1)(d-k+1) k!}\left( \frac{d+1-k} 2\right)_{k/2}
\pold _k \nn\\ & &-\frac{(d-k)}{4(d-1)(k-1)!} \left( \frac{d+2-k}
2 \right) _{(k-1)/2}\pold _{k-1}\nn\\ & &-\frac{2\sqrt{\pi}}
{(d-1)} \sum_{i=1}^{k-1}
\frac{d+i-k}{(k-1-i)!}\left(\frac{d+2-k+i} 2 \right)_{
(k-i-1)/2}\times\label{eq3.24}\\ & &\qquad
\sum_{b=0}^i\frac{x_{i,b}}{\Gamma\left( b +i/2\right)}
\left(\frac{d+1-k+i} 2 \right)_{b} D_{k-1-i}^{d-1} ,\nn \eeq where
$(y)_n =\Gamma (y+n)/\Gamma (y)$ is the Pochhammer symbol.
Eq.(\ref{eq3.24}) exhibits the heat-kernel coefficients as
explicit functions of the dimension $d$ (partly encoded in
$D_\nu^{(n)}$). Clearly, evaluation of eq. (\ref{eq3.24}) is a
simple routine machine matter, because all ingredients can be
found by simple algebraic computer programs.

For later use, the
polynomials up to $a_{3/2}$ are listed in the following,
\beq
\facb\hem _{1/2} &=& -{1\over 4}
\label{dir1}\\
\fac\hem _1 & =& {{d}\over {3}}
\label{dir2}\\
\facb\hem _{3/2} & = &
   {(10-7d) d \over 384 } .
\label{dir3}
\eeq

For Robin boundary conditions one has to make the modifications
outlined above eq.~(\ref{eq2.48}). The results, up to $\hem
_{3/2}$ are listed below, \beq \facb\hem _{1/2}& =& {1\over 4}
\label{rob1}\\ \fac\hem_1 & =& {d \over 3} + 2\beta \label{rob2}\\
\facb\hem_{3/2}& =& { 192 \beta^2 +96 \beta d +d (2+13d) \over
384} \label{rob3} \eeq Further coefficients could be calculated
with ease, f.e. for the first $20$ coefficients the program needs
about $2$ minutes.

\subsection{Determination of the general heat-kernel coefficients}
\label{sec4.1b}
We now compare, one by one, the general form of the coefficients
with our special case evaluation.
The coefficient $a_0$ is,
by normalization,
$$
a_0 (F,P, \bou ) = (4\pi )^{-D/2} F[\cam ].
$$

The next one is $$ a_{1/2} (F,P, \bou)  = \delta (4\pi )^{-d/2}
F[\partial \cam] .$$

For the ball this means $$ a_{1/2} (F,-\De_\cam, \bou)
          = \delta (4\pi )^{-d/2} F(1) |S^d|.$$

Using the relations (\ref{dir1}) and (\ref{rob1}) we can
immediately determine $\delta$, $$ \delta = \left( -\frac 1 4 ^- ,
\frac 1 4 ^+ \right) . $$

The coefficient $a_{1/2}$ is thus given for a general manifold
from the result on the ball (which was clear of course). Passing
on to $a_1$, the general form is $$ a_1( F,P, \bou) = (4\pi
)^{-D/2} 6^{-1} \left\{
 ( 6 FE + FR ) [\cam ] + (b_0 FK +b_1 F_{;m} + b_2 F S ) [\partial \cam]
\right\}
$$

In our special case on the ball, $K_a^b = \delta_a^b$ and thus $$
a_1(F,-\De_\cam, \bou) = (4\pi )^{-D/2} 6^{-1} {\rm vol} (S^d)
\left\{
 b_0 F(1) d + b_1 F_{,r} (1) + b_2 F(1) S \right\}.
$$

Comparing with the results (\ref{dir2}), (\ref{rob2}), one finds
$$ b_0 = 2 , \qquad  \qquad b_2 = 12 . $$

Taking into account also the smeared calculation
\cite{dowk9803094}, in addition one can find \beq b_1 = (3^-, -3^+
).\nn \eeq

Thus, our special case also gives the entire $a_1$ coefficient
without any further information being needed. It is very important
that the calculation can be performed for an arbitrary ball
dimension $D$, and also for a smearing function $F(r)$. This
allows one to just compare polynomials in $d$ with the associated
extrinsic curvature terms in the general expression and simply to
read off the universal constants in this expression.

We continue with the next higher coefficient, with the general
form, \beq a_{3/2} (F,P,\bou) & = &{\delta\over96 (4\pi )^{d/2}}
\bigg(\!
     F\big(c_0 E +c_1 R\! +c_2 R_{mm} +c_3 K^2\! +c_4 K_{ab} K^{ab}
        c_7 SK\! +c_8 S^2\!\big)\nn\\
& &\phantom{\delta (4\pi )^{-d/2} 96^{-1}}
            +F_{;m} (c_5 K +c_9 S) +c_6 F_{;mm} \bigg) .\nn
\eeq

The (smeared) ball calculation immediately gives $7$ of the $10$
unknowns, \beq & &c_3 = (7^-,13^+), \quad  c_4 = (-10 /, 2^+),
\quad c_5 = (30^-, -6^+), \nn\\ & &c_6 = 24, \quad c_7 = 96, \quad
c_8 = 192, \quad  c_9 = -96. \nn \eeq

Next, apply the lemma on product manifolds (\ref{eq3.11})
\cite{bran90-15-245}. For $a_{3/2}$ this means $$ a_{3/2}
(y,P,\bou) = a_{3/2} (y_1,P_1,\bou)
       a_0 ( x_2, P_2) +a_{1/2} (y_1,P_1,\bou) a_1
(x_2, P_2). $$

We will choose $P_1 = -\Delta_1$ and $P_2 = -\Delta_2 + E (x_2)$,
with obvious notation, to obtain

$$ \delta 96^{-1} (c_0 E + c_1 R(\cam _2) ) = \delta 6^{-1} (6 E +
R (\cam _2)), $$ where we used in addition $R( \cam_1 \times
\cam_2 ) = R(\cam _1) + R(\cam _2 )$. This gives

$$ c_0 = 96, \quad c_1 = 16. $$

It is seen that the determination of $a_{3/2}$ is relatively
simple, once the ball result is at hand. The lemma on product
manifolds is also very easily applied and already only one of the
universal constants $c_i$, namely $c_2$, is missing.

The remaining information is obtained using the relations between the heat-
kernel coefficients under conformal rescaling, (\ref{eq3.14}).
Setting to zero the coefficients of all terms in (\ref{eq3.14})
gives several
relations between the universal constants $c_i$. We will need only one of
them. Thus, setting to zero the coefficient of $F_{;mm}$ gives
$$
\frac 1 2 (D-2) c_0 - 2 (D-1) c_1 -(D-1) c_2 -(D-3) c_6 =0
$$
and so $c_2 = -8$ for Dirichlet and Robin boundary conditions.
This completes the calculation of $a_{3/2}$.

It gets clear, already, that the combination of the different
methods is extremely effective in obtaining heat-kernel
coefficients and, as mentioned, $a_{5/2}$ has been obtained
proceeding in this way. For other boundary conditions, the general
form of the heat-kernel coefficients can contain several other
geometrical quantities, as projectors for mixed boundary
conditions (for a discussion of their physical relevance see
\cite{luck91-32-1755}) or tangential vector fields for oblique
boundary conditions. Although the analysis gets partly more
involved, the ideas described seem to apply generally and, as
mentioned, many new results have already been obtained
\cite{avra98-15-281,avra99-200-495,dowk97-14-169,dowk9806168,dowk98pro}.

\section{Casimir and ground state energies}
\setcounter{equation}{0}
\label{sec6}
In this section we apply the techniques developed in Section \ref{sec3.1}
to the calculation of vacuum energies when spherically symmetric
boundaries or external potentials are present. The boundary calculation
is an immediate continuation of Section \ref{sec3.1}. When external
potentials are present, the implicit eigenvalue equation
(\ref{eq2.4}) is replaced by an asymptotic equivalent, see
eq.~(\ref{2.3}), which allows the energy to be expressed by quantum
mechanical scattering dates.

\subsection{Massive scalar field with Dirichlet boundary conditions}
\label{sec6.5}

So let us start to analyse the influence the mass of a field has
on the Casimir energy. We have seen that the Casimir energy is
generally plagued by ambiguities. For a massive field, however,
there is the possibility to separate (at least in principle) the
classical part and the quantum part of the energy. The idea is
that, if the mass of the quantum field tends to infinity, quantum
fluctuations should die out and as a result, in this limit, the
quantum contributions to the Casimir energy should vanish. As we
will see, this provides a unique definition \cite{bord97-56-4896}.
Before we proceed, let us first discuss this in more detail.

To actually impose the condition, the $m\to \infty$ behaviour of $E_{vac}$
is needed. This is easily derived using the heat-kernel expansion. For the
scalar field, the heat-kernel of $(-\De +m^2)$ clearly is
\beq
K(t) = e^{-m^2 t} K_{-\De} (t),\nn
\eeq
with $K_{-\De} (t)$ the heat-kernel of minus the Laplacian on the ball. The
asymptotic expansion for $m\to \infty$ is then found by employing
eq.~(\ref{eq1.6}),
\beq
\zeta (\alpha ) &=& \frac 1 {\G (\alpha )} \int_0^\infty dt 
   \,\,t ^{\alpha -1} e^{-m^2t}
                      K_{-\De} (t) \nn\\
    &\sim &  \frac 1 {\G (\alpha )} \sum_{l=0,1/2,1,...} ^\infty
           a_l (-\Delta ) \frac{\G 
      (\alpha +l-3/2)}{m^{2(\alpha +l-3/2)}} .\label{eq6.4.0}
\eeq

About $\alpha =-1/2$ it reads, \beq \ze (-1/2+s) &=&
-\frac{m^4}{4\sqrt{\pi} } a_0 \left( \frac 1 {s} -\frac 1 2 +\ln
\left[\frac{4\mu^2}{m^2}\right]\right)
-\frac{2m^3}{3}a_{1/2}\nonumber\\ & &+\frac{m^2}{2\sqrt{\pi}} a_1
\left( \frac 1 {s} - 1  +\ln
\left[\frac{4\mu^2}{m^2}\right]\right) + m a_{3/2}\label{n7}\\ &
&-\frac 1 {2\sqrt{\pi}} a_2 \left( \frac 1 {s} - 2 +\ln
\left[\frac{4\mu^2}{m^2}\right]\right)  +\cao (1/m) +\cao (s) .
\label{eq6.4.1} \eeq

Thus, in order to impose the normalization condition \beq
\lim_{m\to\infty} E_{vac}^{ren} =0,\label{eq6.4.2} \eeq this
defines the terms to be subtracted. As is seen, the zeta
functional regularization used leaves the contributions of the
coefficients with half integer index finite in the limit $s\to 0$.
This is a specific feature of this regularization. However, in
other regularizations, as for example the proper time cutoff
\cite{blau88-310-163} or the exponential cutoff
\cite{bene96-11-2871}, these contributions are divergent when the
cutoff is removed. For this reason, and in order to impose the
normalization condition (\ref{eq6.4.2}), we include them among the
"divergent" terms suffering renormalization. By this, we are led
to the following definition of the renormalized Casimir energy,
\beq E_{vac}^{ren} = E_{vac} -E_{vac}^{div} \label{eq6.4.2a} \eeq
with \beq E_{vac}^{div} &=& -\frac{m^4}{8\sqrt{\pi} } a_0 \left(
\frac 1 {s} -\frac 1 2 +\ln \left[\frac{4\mu^2}{m^2}\right]\right)
-\frac{ m^3}{3}a_{1/2}\nonumber\\ & &+\frac{m^2}{4\sqrt{\pi}} a_1
\left( \frac 1 {s} - 1  +\ln
\left[\frac{4\mu^2}{m^2}\right]\right) +\frac 1 2  m a_{3/2}\nn\\
& &-\frac 1 {4\sqrt{\pi}} a_2 \left( \frac 1 {s} - 2 +\ln
\left[\frac{4\mu^2}{m^2}\right]\right). \label{eq6.4.2b} \eeq

To interpret the subtraction in (\ref{eq6.4.2a}) as a
renormalization of "bare" parameters, we need to consider the
following physical system composed of two parts:
\begin{enumerate}
\item A classical system consisting of a spherical
 surface of radius $a$. Its energy reads:
\begin{eqnarray}
E_{class} = p V + \sigma  S + F a +k +\frac{h}{a} ,\label{eq6.4.3}
\end{eqnarray}
where $V=\frac{4}{3}\pi a^3$ and $S=4\pi a^2$ are the volume and
surface, respectively. This energy is determined by the
 parameters $p =$  pressure, $\sigma =$  surface tension,
and $F$, $k$, and $h$, which do not have special names.
\item A quantized field $\hat{\varphi} (x)$ whose classical counterpart obeys
the Klein-Gordon equation
\begin{eqnarray} (\Box +m^2)\varphi (x) =0, \end{eqnarray}
as well as suitable boundary conditions on the surface
ensuring selfadjointness of the corresponding elliptic
operator on perturbations. We choose Dirichlet
boundary conditions as the easiest to handle.
\end{enumerate}

For this system one can consider three models, which will behave
in a different way \cite{bord97-56-4896}. These models consist of
the classical part given by the surface and
\begin{itemize}
\item[ (i)] the quantized field in the interior of the surface,
\item[ (ii)] the quantized field in the exterior of the surface,
\item[ (iii)] the quantized field in both regions together,
\end{itemize}
respectively.
We will give here the analysis for model (i); models (ii) and (iii) can be
treated along the same lines \cite{bord97-56-4896} and only a few comments
will be added.

The \hks needed to impose the normalization (\ref{eq6.4.2}) are
well known (see for instance \cite{kenn78-11-173}, or, alternatively, the
calculation provided in Section \ref{sec3.2}), and read
\beq
a_0 = \frac 1 {6\sqrt{\pi}} a^3 , \quad
a_{1/2} = -\frac 1 4 a^2 , \quad a_1  =
                     \frac 1 {3\sqrt{\pi} } a , \nn\\
a_{3/2}  = -\frac 1 {24} , \quad a_2 = \frac 2
           {315 \sqrt{\pi} a } .\nn
\eeq

Thus, we have five divergent contributions. This remains true for
model (ii); instead, for model (iii), only two divergent terms
remain due to a cancellation of poles which is easily understood
by realizing that the extrinsic curvature seen from the interior
and exterior has opposite sign.

As the physical system we consider, as described, the classical
part (\ref{eq6.4.3}) and the ground state energy of the quantum
field together, and write for the complete energy
\begin{eqnarray}
E=E_{class}+E_{vac}.
\end{eqnarray}

In this context, the renormalization can be achieved by shifting
the parameters in $E_{class}$ by an amount which cancels the
divergent contributions and removes completely the contribution of
the corresponding heat kernel coefficients. In detail we have
\begin{eqnarray}
p &\to &p -   \frac{m^4} {64\pi^2}
\left( \frac 1 {s}
-\frac 1 2 +\ln \left[\frac{4\mu^2}{m^2}\right]\right),\quad
\sigma  \,\to  \, \sigma +\frac{m^3}{48\pi}, \nonumber\\
F  &\to &  F +  \frac{m^2}{12\pi}
\left( \frac 1 {s}
- 1  +\ln \left[\frac{4\mu^2}{m^2}\right]\right), \quad
k\, \to \, k-\frac m {96},
\label{n8}\\
h& \to & h +  \frac 1 {630\pi}
\left( \frac 1 {s}
- 2  +\ln \left[\frac{4\mu^2}{m^2}\right]\right).\nonumber
\end{eqnarray}

After the subtraction of these contributions from $E_{vac}$ we
denote it by $E_{vac}^{ren}$, see eq.~(\ref{eq6.4.2a}), and the
complete energy becomes
\begin{eqnarray}
E=E_{class}+E_{vac}^{ren},
\end{eqnarray}
with the bare parameters in $E_{class}$ replaced by the renormalized
ones using eq.~{\ref{n8}).
In our renormalization scheme, we have defined a unique renormalized
groundstate energy $E_{vac}^{ren}$.

Now, having discussed in detail the subtraction procedure and the structure of
divergences, let us come to a full evaluation of $E_{vac}$. The equations
of Section \ref{sec3.1} serve as a starting point.
Choosing $N=3$ in eq.~(\ref{eq2.12}),
$Z(s)$ is finite at $s=-1/2$ and can be used for the
numerical evaluation of $E_{vac}$ for (in principle) arbitrary mass $m$.
For $A_i (s)$, $i=-1,0,1,2,3$, a representation valid for $|ma|<1/2$ has
been given, see eqs.~(\ref{eq2.24}), (\ref{eq2.25}) and
(\ref{eq2.28}), but here we need a representation of $A_i (s)$ beyond
that range. This is a purely technical complication explained
in Appendix \ref{anhda}. The final result for $A_{-1} (s)$ and
$A_0 (s)$ about $s=-1/2$ is given in eqs.~(\ref{anhdaa}) and
(\ref{anhdab}). The remaining $A_i (s)$ have the form, see
eq.~(\ref{eq2.27}),
\beq
A_i (s) &=& -\frac {2m^{-2s} }{\G (s)} \sum_{c=0}^i \frac
       {x_{i,c} }{(ma)^{i+2c}} \frac{\G (s+c+i/2)}{\G (c+i/2)}
           f(s; 1+2c; c+i/2 ) ,\label{eq6.47}
\eeq
with
\beq
f(s;c;b) =\sum_{\nu = 1/2,3/2,\ldots} ^{\infty}
\nu ^c \left( 1+ \left( \frac{\nu}
{ma } \right ) ^2 \right) ^{-s-b}. \label{fab}
\eeq

The remaining task is to calculate $f(s;c;b)$ for the relevant
values of $c$ and $b$ about $s=-1/2$. This is a systematic
calculation sketched also in Appendix \ref{anhda}. It is slightly
simplified by realizing the recurrence \beq f(s;c;b) = (ma)^2
\left[ f(s;c-2;b-1)- f(s;c-2;b)\right]. \label{recur} \eeq

In summary, all analytic expressions needed for the numerical
evaluation have been provided.

The result for the Casimir energy for the interior region of the ball is
shown in Fig.~\ref{fig5} for $a=1$ as a function of $m$.

\begin{figure}[ht]
\setlength{\unitlength}{1cm}
\begin{center}
\begin{picture}(13,10)
\epsfig{figure=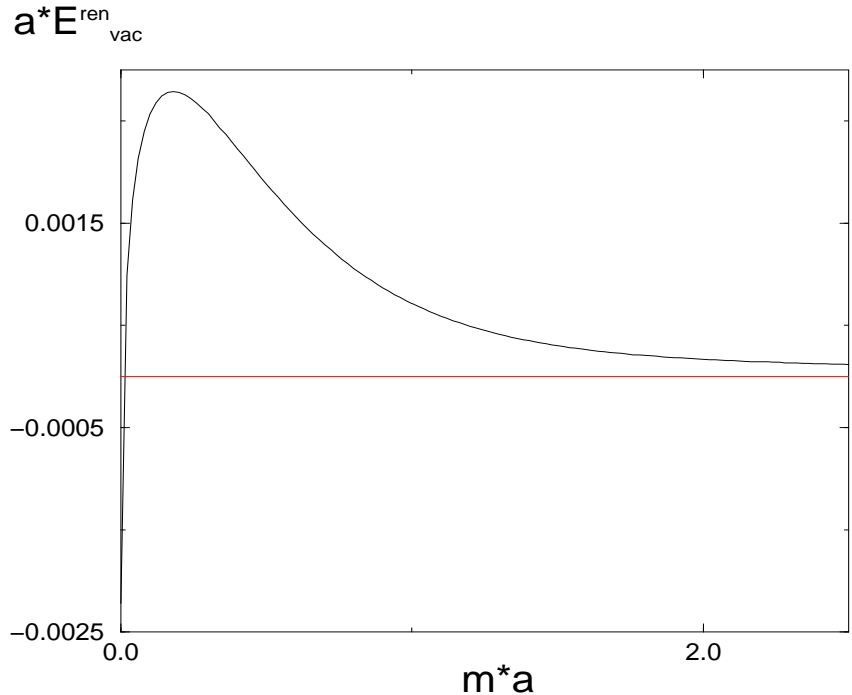,height=10cm,width=13cm}
\end{picture}

\caption{\label{fig5}Plot of the renormalized vacuum energy
$E^{ren}_{vac}$ measured in units of the inverse of the radius.}

\end{center}

\end{figure}

For very small values of the argument, $ma$, the function takes
negative values, whereas for larger values it is positive and
tends to zero for $(ma)\to\infty$. Thus, as a function of the
mass, attractive as well as repulsive forces are possible and the
influence of the mass is by no means negligeable. This is
different from the example of parallel plates
\cite{ambj83-147-1,plun86-134-87}, where for large mass the
influence of the mass is exponentially damped away and is of a
very short range. The origin of this different behaviour is
intimately connected with the extrinsic curvature being
non-vanishing or vanishing. This is already seen from the
heat-kernel expansion (\ref{eq6.4.0}), where the leading
contribution to $E_{vac}^{ren}$ is proportional to $a_{5/2} (-\De
)/m$. For the ball, this coefficient is non-vanishing; however,
for parallel plates this one as well as all the following
coefficients are vanishing, what explains the very different
behaviour of the renormalized Casimir energy as a function of the
mass for this two configurations.

\subsection{Scalar field in the presence of an external background field}
\label{sec7.1}

In the above described situation, "empty space" is supplemented by
boundary conditions. However, clearly (perfect) boundaries are an
idealization and no boundary made of matter can be perfectly
smooth. Seen as modeling some distribution of matter which
interacts with the quantum field, it might be more justified to
introduce an external potential and to quantize a theory within
this potential. This will change the quantum modes and the
associated spectrum and, as a result, the energy of the vacuum.
For certain types of external confinement this situation has been
called Casimir effect with soft or semihard boundaries
\cite{acto95-52-3581,acto95u}. As mentioned, apart from this type
of setting, in many situations the potential is provided by
classical solutions to nonlinear field equations.

Again, we will start the consideration by describing our concrete
model and its renormalization, introducing also various notations
used in the following. We will consider the Lagrangian
\begin{eqnarray}
L=\frac 1 2 \Phi (\Box -M^2 -\lambda \Phi ^2 )\Phi
+\frac 1 2 \varphi (\Box -m^2 -\lambda ' \Phi^2 )\varphi .\label{22.1}
\end{eqnarray}

Here, the field $\Phi$ is a classical background field. By means
of
\begin{eqnarray}
V(x) =\lambda ' \Phi ^2  \label{22.2}
\end{eqnarray}
it defines the potential in (\ref{22.1}) for the field $\varphi
(x)$, which should be quantized in the background of $V(x)$. As
explained below, the embedding into this external system is
necessary in order to guarantee the renormalizability of the
ground state energy. Actually, this is clear already from the
beginning, because the external system needs to comprise of the
counterterms of a $(\lambda \varphi^4$)-theory.

The complete energy
\begin{eqnarray}
E[\Phi ] =E_{class} [\Phi ] +E_{vac} [\Phi ]\label{E}
\end{eqnarray}
of the system consists, as before, of the classical part and the
contributions resulting from the ground state energy of the
quantum field $\varphi$ in the background of the field $\Phi$. The
classical part reads
\begin{eqnarray}
E_{class}[\Phi ]=\frac 1 2 V_g +\frac 1 2 M^2 V_1 +\lambda V_2, \label{22.3}
\end{eqnarray}
with the definitions $V_g =\int d^3x (\nabla \Phi )^2$, $V_1 =\int
d^3x \Phi^2$ and $V_2 =\int d^3x \Phi^4$. Here $M^2$ and $\lambda$
are the bare mass, respectively coupling constant, which need
renormalization. Continuing as in Section \ref{sec6}, for the
ground state energy one defines \cite{blau88-310-163}
\begin{eqnarray}
E_{vac}[\Phi ] =\frac 1 2 \sum_{(n)} (\lambda_{(n)} ^2 +m^2 )
 ^{1/2-s} \mu^{2s}
,\label{22.4}
\end{eqnarray}
where $\mu$ is the arbitrary mass parameter and $s$ is the
regularization parameter, which has to be put to zero after
renormalization. Furthermore, $\lambda_{(n)}$ are the eigenvalues
of the corresponding Laplace equation
\begin{eqnarray}
(-\Delta +V(x))\phi_{(n)}(x) =\lambda _{(n)}^2 \phi_{(n)}(x) .\label{22.5}
\end{eqnarray}

For the moment, we assume the space to be a large ball of radius
$R$ as an intermediate step to have discrete eigenvalues and,
thus, a discrete multiindex $(n)$ and, in addition, to avoid pure
volume divergences. For simplicity we impose Dirichlet \bcs and
explain below under which conditions on the potential $V(x)$ the
final result does not receive boundary contributions.

Expressing the ground state energy (\ref{22.4}) in terms of
the zeta function
\begin{eqnarray}
\zeta_V (s) = \sum_{(n)} (\lambda_{(n)} ^2 +m^2 )^{-s}\label{zetascat}
\end{eqnarray}
of the wave operator with potential $V(x)$ as defined in (\ref{22.5}),
one has
\begin{eqnarray}
E_{vac } [\Phi ] =\frac 1 2 \zeta_V (s-1/2) \mu^{2s}.\label{groundzeta}
\end{eqnarray}

The residue of $\zeta_V (-1/2)$ is determined by the $a_2$
heat-kernel coefficient associated with the spectrum
$\lambda_{(n)}^2 +m^2$, and the classical energy (\ref{22.3})
provides another example for our previous viewpoint, that it
should consist of the terms contained in $a_2$. In order to fix
the finite renormalizations, we impose again the normalization
condition \beq \lim_{m\to\infty} E_{vac}^{ren} [\Phi ] =0
,\label{grundnorm} \eeq which is implemented along the lines
leading to eq.~(\ref{eq6.4.2b}). For convenience we state it
again,
\begin{eqnarray}
E_{vac} ^{div} [\Phi ] &=& -\frac{m^4}{
8\sqrt{\pi}}\left(\frac 1 {s}
+\ln\frac
{4\mu^2}{m^2} -\frac 1 2 \right) a_0 -\frac{m^3} 3 a_{1/2} \nonumber\\
& &+\frac{m^2}{4\sqrt{\pi}}\left(\frac 1 {s}
+\ln\frac{4\mu^2}{m^2} -1\right) a_1 +\frac 1 2 a_{3/2}
\label{endiv}\\
& &-\frac 1 {4\sqrt{\pi}} \left(
\frac 1 {s} +\ln \frac{4\mu^2}{m^2}-2\right) a_2,\nonumber
\end{eqnarray}
with $a_{i} , i=0, 1/2,...,2$ the heat-kernel coefficients of the
Laplace equation (\ref{22.5}). This is a good moment to consider
the limiting behaviour for $R\to\infty$ of the boundary terms.
Before actually performing the limit $R\to\infty$ we must clearly
state what the ground state energy in the given context is meant
to be. It is a quantity that describes the influence of the
background potential $V(x)$ on the vacuum energy and, as such, is
to be compared with the field free case $V(x)=0$. Thus, from
definition (\ref{22.4}), we will subtract the Casimir energy of a
free field inside a large ball of radius $R$ in order to normalize
$E_{vac} [\Phi =0 ]=0$. In eq.~(\ref{endiv}), this subtraction
cancels the contributions of $a_0$ and $a_{1/2}$, in $a_1$ it
cancels the potential independent boundary terms. In $a_{3/2}$ the
term proportional to $V(R) R^2$ survives, see eq.~(\ref{eq3.8}), and,
in the limit $R\to\infty$ vanishes only if $V(r) \sim r^{-2-\ep}$,
$\ep > 0$, for $r\to\infty$. As is easily seen by dimensional
arguments, the boundary contributions of the higher coefficients
depending on $V(r)$ will then vanish too.

In summary, under the assumption that $V(r) \sim r^{-2-\ep}$ for
$r\to\infty$, in the limit $R\to\infty$ no boundary contributions
will appear and the normalization condition (\ref{grundnorm}) is
achieved by a renormalization of the mass $M$ of the background
field,
\begin{eqnarray}
M^2 & \to & M^2 +\frac{\lambda 'm^2}{16 \pi^2 } \left( -\frac 1 {s} +1 +\ln
\frac{m^2}{4\mu^2}\right),\label{22.7}
\end{eqnarray}
and the coupling constant $\lambda$ by
\begin{eqnarray}
\lambda & \to & \lambda +\frac{{\lambda '} ^2}{64\pi^2} \left(
-\frac 1 {s} +2 +\ln\frac{m^2}{4\mu^2}\right).\label{22.8}
\end{eqnarray}

This is an immediate result of the volume contributions to the
heat-kernel coefficients, $a_1 =-(4\pi)^{-3/2}\int d^3x V(x)$ and
$a_2 =(1/2) (4\pi)^{-3/2}\int d^3x V^2(x)$. The kinetic term $V_g$
in $E [\Phi ]$ suffers no renormalization.

By
defining
\begin{eqnarray}
E_{vac}^{ren} =E_{vac} [\Phi ] -E_{vac}^{div}\label{renen}
\end{eqnarray}
one obtains the finite groundstate energy, which is normalized in a way that
the functional dependence on $\Phi ^2$ present in the classical energy is now
absent in the quantum corrections $E_{vac}^{ren}$.

Let us note that this is just the well known general renormalization scheme
written down here explicitly in the notations needed in our case.

For the calculation of $E_{vac} [\Phi ]$ let us take further advantage
of the background field $\Phi (r)$ being spherically symmetric.
The multiindex $(n)\to n,l,m$ consists of the main quantum number $n$, the
angular momentum number $l$ and the magnetic quantum number $m$.
In polar coordinates the ansatz for a solution of the wave equation (\ref{22.5})
reads
\begin{eqnarray}
\phi _{(n)} (x) =\frac 1 r \phi_{n,l} (r) Y_{lm}(\theta , \varphi )\label{ansatz}
\end{eqnarray}
where the radial wave equation takes the form
\begin{eqnarray}
\left[\frac{d^2 }{dr^2} -\frac{l(l+1)}{r^2} -V (r)
+\lambda_{n,l}^2 \right]
\phi_{n,l} (r) =0. \label{2.1}
\end{eqnarray}

Now we use the standard scattering theory within $r\in [0,\infty)$
and take the momentum $p$ instead of the discrete $\lambda_{n,l}$.
Let $\phi_{p,l} (r)$
 be  the so called
regular solution which is defined so as to have the same behaviour
at $r\to 0$ as the solution without potential
\begin{eqnarray}
\phi_{p,l} (r) &\sim& j_l (pr)\label{regular}\\[-5pt]
&{\stackrel {r \to 0}{}}&\nonumber
\end{eqnarray}
with $j_l $ the spherical Bessel function \cite{tayl72b}. This
regular solution defines the Jost function $f_l$ through its
asymptotics as $r\to\infty$,
\begin{eqnarray}
\phi_{l,p}(r) &\sim& \frac i 2 \left[ f_l(p) \hat h _l^- (pr) -
        f_l^*(p) \hat h _l^+ (pr) \right],\label{2.2}\\[-5pt]
&{\stackrel{r\to\infty}{}}&\nonumber
\end{eqnarray}
where $\hat h _l^- (pr)$
and $\hat h _l^+ (pr)$ are the Riccati-Hankel functions \cite{tayl72b}.
The analytic properties of the Jost function $f_l (p)$ strongly depend
on the properties of the potential $V(r)$. If in addition to
$V(r) \sim r^{-2-\ep} $ for $r\to\infty$, we impose $V(r) \sim
r^{-2+\ep} $ for $\ep \to 0$ and continuity of $V(r) $ in $0<r<\infty$
(except maybe at a finite number of finite discontinuities), one may show
that the Jost function is an analytic function of $p$ for $\Im p > 0$.
It vanishes in the finite
set of points $p=i\kappa_{n,l}$ of the positive imaginary
half axis, corresponding to the bound states with energy $-\kappa_{n,l}^2$.

Now we use the Jost function to transform the frequency sum in
eq.~(\ref{zetascat}) in a contour integral. Let us assume that the
support of the potential is contained in the cavity of radius $R$.
Then the above eq.~(\ref{2.2}) gets exact at $r=R$ and may be interpreted as
an implicit equation for the eigenvalues $p=\lambda_{n,l}$.
Choosing Dirichlet boundary conditions at $r=R$, $\phi_{p,l}(R) =0$, it reads
explicitly,
\begin{eqnarray}
 f_l(p) \hat h _l^- (pR) - f_l^*(p) \hat h _l^+ (pR) =0.\label{2.3}
\end{eqnarray}

As already mentioned, ultimately we are interested in the limit
$R\to \infty$ and in that limit the results will not receive
boundary contributions, once we assume that $V(r) \sim
r^{-2-\epsilon}$ for $r\to \infty$.

Let us now consider the ground state energy associated with the eigenvalues
determined by (\ref{2.3}). As described in detail in Section \ref{sec3.1},
we represent the frequency sum
in (\ref{22.4}) by a contour integral.
Using eq.~(\ref{2.3}), one immediately
finds
\begin{eqnarray}
E_{vac} [\Phi ] &=& \mu^{2s}\sum_{l=0}^{\infty}(l+1/2)
\int\limits_{\gamma}\frac{dp}{2\pi i}\,\,
(p^2+m^2)^{1/2-s}\frac{\partial}{\partial p}
               \ln
\left[
\frac{f_l(p) \hat h _l^- (pR) - f_l^*(p) \hat h _l^+ (pR)}
    {h _l^- (pR) -\hat h _l^+ (pR)}\right] \nonumber\\
& &+\mu^{2s}\sum_{l=0}^{\infty}(l+1/2) \sum_n (m^2-\kappa_{n,l}^2)^{1/2-s},\label{2.4}
\end{eqnarray}
with $-\kappa_{n,l}^2$ the energy eigenvalues of the bound states
with given orbital momentum $l$. The denominator in the logarithm
provides the subtraction of the Minkowski space contribution,
where $f_l (p)=1$. The contour $\gamma$ is chosen counterclockwise
enclosing all real solutions of eq.~(\ref{2.3}) on the positive
real axis. In the limit of the infinite space the negative
eigenvalues become the usual boundstates and the $\lambda_{n,l} >
0$ turn into the scattering states.

For the calculation of (\ref{2.4}), as the next step, one deforms
the contour $\gamma$ to the imaginary axis. A contour coming from
$i \infty +\epsilon $, crossing the imaginary axis at some
positive value smaller than the smallest $\kappa _n$ and going to
$i\infty -\epsilon$ results first. Shifting the contour over the
bound state values $\kappa_n $, which are the zeroes of the Jost
function on the imaginary axis, the bound state contributions in
eq.~(\ref{2.4}) are cancelled and in the limit $R\to \infty$ one
finds
\begin{eqnarray}
E_{vac} [\Phi ]  =-{\cos \pi s\over \pi}\mu^{2s} \sum_{l=0}^{\infty}(l+1/2)
\int\limits_{m}^{\infty}dk\,\, [k^2-m^2]^{\frac{1}{2}-s}~\frac{\partial}{\partial k}\ln  f_l (ik).\label{2.5}
\end{eqnarray}

For details of the contour deformation  see Fig.~\ref{fig10}.

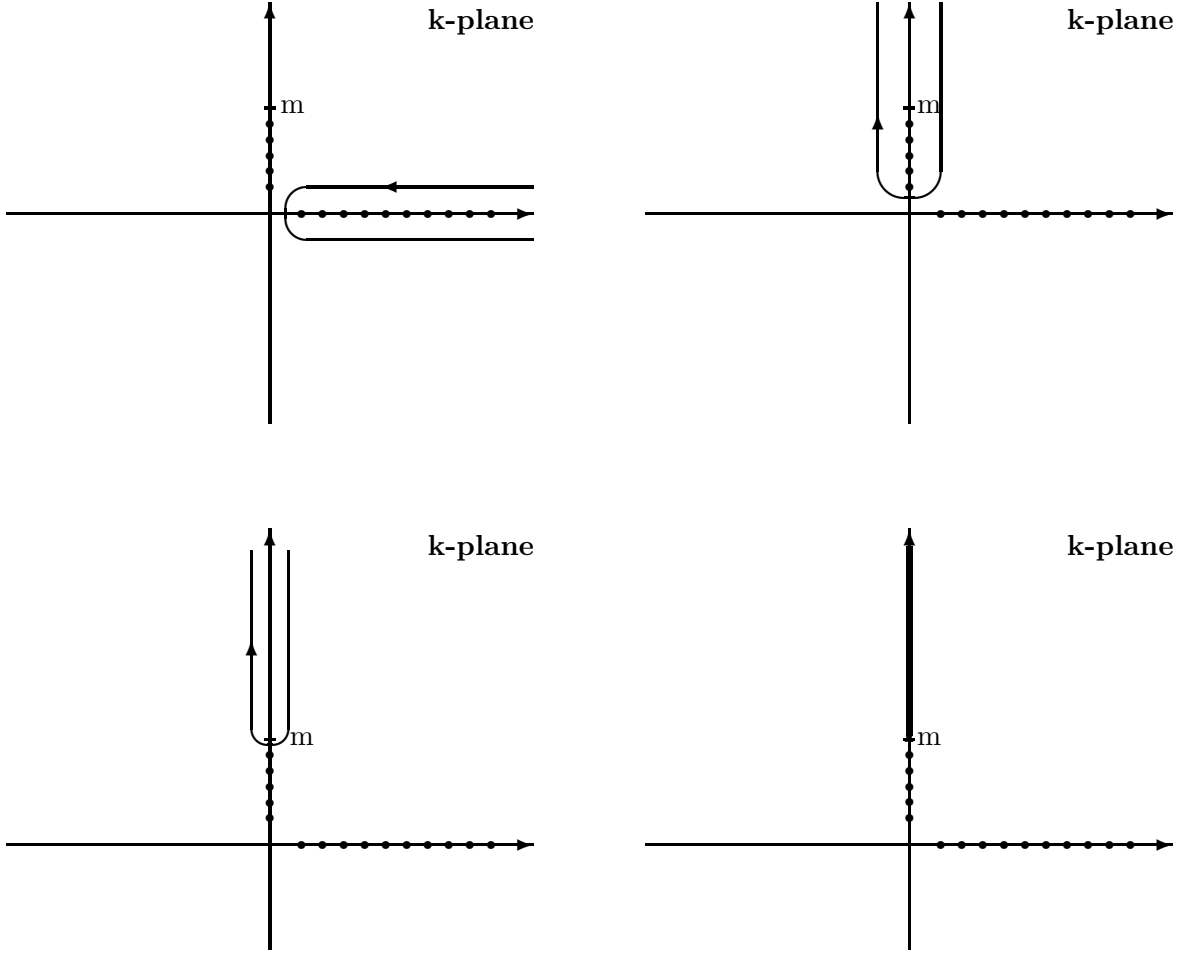
\begin{figure}[ht]
\setlength{\unitlength}{1cm}

\begin{center}

\begin{picture}(20,14)(0,-7.5)

\put(0,0){\setlength{\unitlength}{.7cm}
\begin{picture}(10,7.5)
\thicklines

\put(0,4){\vector(1,0){10}}
\put(5.0,0){\vector(0,1){8}}
\put(10.0,4){\oval(9.4,1)[l]}  \put(7.5,4.5){\vector(-1,0){.4}}
\multiput(5.6,4)(.4,0){10}{\circle*{.15}}
\multiput(5.0,4.5)(0,.3){5}{\circle*{.15}}
\put(4.9,6){\line(1,0){.2}}
\put(5.2,5.9){m}
\put(8.0,7.5){{\bf k-plane}}

\end{picture}}

\put(8.50,0){\setlength{\unitlength}{.7cm}
\begin{picture}(10,7.5)
\thicklines

\put(0,4){\vector(1,0){10}}
\put(5.0,0){\vector(0,1){8}}
\put(5,8){\oval(1.2,7.4)[b]}
\put(4.4,5.5){\vector(0,1){.4}}
\multiput(5.6,4)(.4,0){10}{\circle*{.15}}
\multiput(5.0,4.5)(0,.3){5}{\circle*{.15}}
\put(4.9,6){\line(1,0){.2}}
\put(5.15,5.9){m}
\put(8.0,7.5){{\bf k-plane}}

\end{picture}}

\put(0.0,-7){\setlength{\unitlength}{.7cm}
\begin{picture}(10,7.5)
\thicklines

\put(0,2){\vector(1,0){10}}
\put(5.0,0){\vector(0,1){8}}
\put(5,7.6){\oval(0.7,7.4)[b]}
\put(4.65,5.5){\vector(0,1){.4}}
\multiput(5.6,2)(.4,0){10}{\circle*{.15}}
\multiput(5.0,2.5)(0,.3){5}{\circle*{.15}}
\put(4.9,4){\line(1,0){.2}}
\put(5.3775,3.9){m}
\put(8.0,7.5){{\bf k-plane}}
\end{picture}}

\put(8.5,-7){\setlength{\unitlength}{.7cm}
\begin{picture}(10,7.5)
\thicklines

\put(0,2){\vector(1,0){10}}
\put(5.0,0){\vector(0,1){8}}
\put(5,7.675){\oval(0.075,7.4)[b]}
\multiput(5.6,2)(.4,0){10}{\circle*{.15}}
\multiput(5.0,2.5)(0,.3){5}{\circle*{.15}}
\put(4.9,4){\line(1,0){.2}}
\put(5.15,3.9){m}
\put(8.0,7.5){{\bf k-plane}}
\end{picture}}

\end{picture}

\caption{\label{fig10}Deformation of the contour $\gamma$}

\end{center}

\end{figure}
In the first step of the deformation one makes use of the following
properties \cite{tayl72b},
\beq
f_l (-p) &=& f_l^*(p) ,\nn\\
\hat h _l^{\pm} (-z) &=& (-1)^l \hat h _l^{\mp} (z)  .\nn
\eeq

This is the representation of the ground state energy (and by means of
(\ref{groundzeta}) as well of the zeta function)
in terms of the Jost function,
which is the starting point of our following analysis. It has the nice
property, that the dependence on the bound states is not present explicitly,
being however contained in the Jost function by its properties on the positive
imaginary axis.
It is connected with the more conventional representations
by means of the analytic properties of the Jost function. Expressing
them by the dispersion relation
\begin{eqnarray}
f_l (ik) =\prod_n\left(1-\frac{\kappa_{n,l}^2}{k^2}\right) \exp\left(
-\frac 2 {\pi}\int_0^{\infty}\frac{dq\,\, q}{q^2 +k^2} \delta_l (q)\right),
\nonumber
\end{eqnarray}
where $\delta_l (q)$ is the scattering phase, we obtain from (\ref{2.5})
\begin{eqnarray} E_{vac} [\Phi ]
=\mu^{2s}\sum_{l=0}^\infty \left(l+\frac{1}{2}\right) \left\{-
\sum_n \left(
m^{1-2s}-\sqrt{m^2-\kappa_{n,l}^2}^{\,1-2s}\right)\right.\label{Eun}\\
\left.-\frac{1-2s}{\pi}\int_0^{\infty}dq\,\,
{q\over\sqrt{q^2+m^2}^{1-2s}}\,\delta_l
(q)\right\},\nonumber\end{eqnarray} which gives the expression of
the ground state energy through the  scattering phase. From here
one can pass to the representation through the mode density by
integrating by parts.

Note that this representation can be obtained also directly from
(\ref{2.4}) in the limit $R\to \infty$ by deforming the contour
$\gamma$.

Let us add a discussion on the sign of the ground state energy. In
(\ref{Eun}) the first contribution results from the bound states
and is completely negative. The second contribution, which
contains the scattering phase $\delta_l(q)$, is positive
(negative) for an attractive (repulsive) potential, i.e., for
$V(r)<0$ ($V(r)>0$) for all $r$ \cite{tayl72b}. So the regularized
(still not renormalized) ground state energy $E_{vac} [\Phi ]$
(\ref{Eun}) is positive for a potential which is repulsive for all
$r$ (there are no bound states in this case) and it is negative
for a potential which is attractive for all $r$. Now, if we
perform the renormalization in accordance with (\ref{22.7}) and
(\ref{22.8}), we obtain
\begin{eqnarray}
E^{ren}_{vac}=E_{vac} [\Phi ]+{m^2\over
8\pi}\left(\frac{1}{s}+\log\frac{4\mu^2}{m^2}-1\right)\int_0^\infty{\rm
d}r\,r^2\,V(r)\nonumber \\ +{1\over
16\pi}\left(\frac{1}{s}+\log\frac{4\mu^2}{m^2}-2\right)\int_0^\infty{\rm
d}r\,r^2\,V(r)^2\,.\label{Eren}\end{eqnarray}

This expression is finite for $s\to 0$, i.e., when removing the
regularization. But due to the subtracted terms there is no longer
any definite result on the sign. Note that this is in contrast to
the case of a one-dimensional potential, where it had been
possible to express the subtracted terms through the scattering
phase \cite{bord95-28-755}.

Let us continue with a detailed analysis of the ground state
energy, eq.~(\ref{2.5}). As we have easily seen using heat-kernel
techniques, the non-renormalized vacuum energy $E_{vac} [\Phi ]$
contains divergencies in $s=0$, see eq.~(\ref{endiv}), which are
removed by the renormalization prescription given in
eqs.~(\ref{22.7}) -- (\ref{renen}), resp. (\ref{Eren}). The poles
present are by no means obvious in the representation (\ref{2.5})
of $E_{vac} [\Phi]$. However, in order to actually perform the
renormalization, eq.~(\ref{renen}), it is necessary to represent
the groundstate energy eq.~(\ref{2.5}) in a form which makes the
explicit subtraction of the divergencies visible. This will be our
first task.

As is known from general zeta function theory as well as one sees
from simply counting
the large momentum behaviour, the representation
eq.~(\ref{2.5}) of $E_{vac} [\Phi ]$ will be convergent for $\,\mbox{Re}\, s >
  2$.
However, for the calculation of the ground state energy we need
the value of eq.~(\ref{2.5}) at $s=0$; thus, an analytical
continuation to the left has to be constructed. The basic idea is
the same as before: adding and subtracting the leading uniform
asymptotics of the integrand in eq.~(\ref{2.5})
\cite{bord96-53-5753}. Let
\begin{eqnarray}
E_{vac} [\Phi ]=E_f+E_{as}, \label{3.3a}
\end{eqnarray}
where
\begin{eqnarray}
 E_f =-\frac{\cos (\pi s)}{\pi} \mu^{2s}
\sum_{l=0}^{\infty}(l+1/2) \int\limits_{m}^{\infty}dk\,\, [k^2-m^2]^{\frac{1}{2}-s}\frac{\partial}{\partial k} [\ln f_l (ik) - \ln f_l ^{asym} (ik) ]
\label{ns}
\end{eqnarray}
and
\begin{eqnarray}
 E_{as} =-\frac{\cos (\pi s)}{\pi} \mu^{2s} \sum_{l=0}^{\infty}(l+1/2) \int\limits_{m}^{\infty}dk\,\, [k^2-m^2]^{\frac{1}{2}-s}\frac{\partial}{\partial k} \ln f_l ^{asym} (ik).\label{as}
\end{eqnarray}

As described in detail, the idea is that as many asymptotic terms
are subtracted as necessary  to take $s=0$ in the integrand of
$E_f$. This term will then (in general) be evaluated numerically.

In $E_{as}$ the analytic continuation to $s=0$ can be done
explicitly, showing that the pole contributions cancel when
subtracting $E_{vac}^{div}[\Phi ]$, eq.~(\ref{endiv}).

The first task, thus, is to obtain the asymptotics of the Jost
functions. Contrary to the case with boundary conditions, the
asymptotics cannot be taken just from tables, but need to be
calculated itself. Fortunately, this may be done by using the
integral equation (Lippmann-Schwinger equation) known from
scattering theory \cite{tayl72b}. For the Jost function one has
($\nu \equiv l+1/2$)
\begin{eqnarray}
f_l (ik) =1+\int\limits_0^{\infty} dr\,\,r\,V(r) \phi_{l,ik} (r)
   K_{\nu} (kr),\label{3.1}
\end{eqnarray}
with the regular solution given by the integral equation
\begin{eqnarray}
\phi_{l,ik} (r) &=& I_{\nu} (kr) +
\int\limits_0^r dr'\,\, r'\,\,[ I_{\nu } (kr) K_{\nu} (kr')-
 I_{\nu} (kr') K_{\nu} (kr)] V(r') \phi_{l,ik}(r'). \label{3.2}
\end{eqnarray}

General zeta function theory tells us that the divergence at $s=0$
contains at most terms of order $V^2$. Thus, one might expand $\ln
f_l (ik)$ in powers of $V$ and take into account only the
asymptotics of terms up to $ {\cal O} (V^2)$. The expansion in
powers of $V$ is easily obtained. Using eqs.~(\ref{3.1}) and
(\ref{3.2}) one finds
\begin{eqnarray}
\ln f_l (ik) &=& \int\limits_0^{\infty}dr\,\, r V(r)
K_{\nu} (kr) I_{\nu} (kr)\nonumber\\
& &-\int\limits_0^{\infty}dr\,\, r V(r)
K_{\nu}^2 (kr) \int\limits_0^rdr'\,\, r'
 V(r') I_{\nu}^2 (kr')\nonumber\\
& &+ {\cal O} (V^3). \label{3.3}
\end{eqnarray}

Now, the uniform asymptotics for $l\to \infty$ of $\ln f_l (ik)$
is essentially reduced to the well known uniform asymptotics of
the modified Bessel functions $K_{\nu}$ and $I_{\nu}$, (9.7.7) and
(9.7.8) in \cite{abra70b}. With the notation
$t=1/\sqrt{1+(kr/\nu)^2}$ and $\eta (k) =\sqrt{1+(kr/\nu)^2}+\ln
[(kr/\nu) /(1+\sqrt{1+(kr/\nu)^2})]$, one finds for
$\nu\to\infty$, $k\to\infty$ with $k/\nu $ fixed,
\begin{eqnarray}
I_{\nu} (kr   ) K_{\nu} (kr ) &\sim &
\frac 1 {2\nu t} +\frac{t^3}{16\nu^3}\left( 1-6 t^2
+5 t^4 \right) + {\cal O} (1/\nu^4 )\nonumber\\
I_{\nu} (k r' ) K_{\nu} (kr)& \sim&
\frac 1 {2\nu} \frac{e^{-\nu (\eta (k) -\eta (kr'/r))}}
{(1+(kr/\nu)^2)^{1/4} (1+(kr'/\nu)^2)^{1/4}}
\left[ 1+ {\cal O} (1/\nu )\right].\nonumber
\end{eqnarray}

Using these terms in the rhs of eq.~(\ref{3.3}) we define
\begin{eqnarray}
\ln f_l^{asym}(ik)&=&
 \frac 1 {2\nu} \int\limits_0^{\infty}dr\,\,
\frac{r\,V(r)}{\left[1+\left(\frac {kr}
{\nu}\right)^2 \right]^{1/2}} \nonumber\\
& &+\frac 1 {16\nu^3}\int\limits_0^{\infty}dr\,\,
\frac{r\,V(r)}{ \left[1+\left(\frac {kr}
{\nu}\right)^2 \right] ^{3/2}}\left[1 -
 \frac 6
  {\left[1+\left(\frac {kr} {\nu}\right)^2 \right] }
+ \frac 5  {\left[1+\left(\frac {kr}
{\nu}\right)^2 \right]^{2}}\right]\nonumber\\
& & -\frac 1 {8\nu^3}\int\limits_0^{\infty}dr\,\,\frac{r^3\, V^2 (r)}
{ \left[1+\left(\frac {kr} {\nu}\right)^2 \right]^{3/2}}.
\label{3.3b}
\end{eqnarray}

Thereby the $r'$-integration in the term quadratic in $V$ has been
performed by the saddlepoint method using the monotony of $\eta
(k)$. Now, by means of (\ref{3.3b})  the limit $s\to 0$ can be
performed in eq. (\ref{ns}) and we obtain
\begin{eqnarray}
 E_f =-\frac{1}{\pi} \sum_{l=0}^{\infty}(l+1/2) {\int\limits_{m}^{\infty}dk\,\, \sqrt{k^2-m^2}}
\frac{\partial}{\partial k}\left(\ln f_l (ik) -\ln f_l^{asym} (ik)
\right) ,
\label {efin}
\end{eqnarray}
a form which is suited for a numerical evaluation.

For
$E_{as}$ at $s=0$ one might explicitly find
the analytical
continuation. First of all, the $k$-integrals may be done
using
\begin{eqnarray}
{\int\limits_{m}^{\infty}dk\,\, [k^2-m^2]^{\frac{1}{2}-s}
\frac{\partial}{\partial k}}\left[1+\left(\frac {kr} {\nu}\right)^2 \right] ^{-\frac{n}{2}}=
 -\frac{\Gamma (s+\frac{n-1}{2})
\Gamma (\frac{3}{2}-s)}{ \Gamma (n/2)}
 \frac{\left(\frac{\nu}{mr}\right)^n
m^{1-2s}}{\left(1+
\left(\frac{\nu}{mr}\right)^2\right)^{s+\frac{n-1}{2}}}.\nn
\end{eqnarray}

This naturally leads to the functions encountered already in the
case of boundary conditions,
\begin{eqnarray}
f(s;c;b) =\sum_{\nu = 1/2,3/2,\ldots} ^{\infty}
\nu ^c \left( 1+ \left( \frac{\nu}
{mr } \right ) ^2 \right) ^{-s-b} . \label{funf}
\end{eqnarray}

In terms of these functions, for $E_{as} $ we obtain \beq E_{as}
[\Phi ]&=& -\frac{\G (s) }{2\sqrt{\pi} \G (s-1/2) } \left(
\frac{\mu }{m}\right)^{2s} \int_0^\infty dr \,\,
          V(r) f(s-1/2;1;1/2) \nn\\
& &+ \frac{\G (s+1)}{4\sqrt{\pi} m^2 \G (s-1/2)}
\left( \frac{\mu }{m}\right)^{2s} \int_0^\infty dr \,\,
 \left[ V^2 (r) -\frac{V(r) }{2r ^2} \right] f(s-1/2,1,3/2) \nn\\
& & +\frac{\G (s+2)}{2\sqrt{\pi}m^4 \G (s-1/2) } \left( \frac{\mu
}{m}\right)^{2s} \int_0^\infty dr \,\, \frac{V(r)}{r^4}
f(s-1/2;3;5/2) \nn\\ & &-\frac{\G (s+3) }{6\sqrt{\pi}m^6 \G
(s-1/2)} \left( \frac{\mu }{m}\right)^{2s} \int_0^\infty dr \,\,
\frac{V(r)}{r^6} f(s-1/2;5;7/2) .\nn \eeq

The relevant expansions about $s=0$ of the $f(s;c;b)$ are found in
Appendix \ref{anhda}. All divergences are made explicit, and \beq
E_{as}^{ren} [\Phi ] = E_{as} [\Phi ] -E_{vac} ^{div} [\Phi ] \nn
\eeq takes the compact form \cite{bord99u}
\begin{eqnarray}
E_{as}^{ren}[\Phi ] &=&
-\frac 1 {8\pi} \int_0^\infty dr \,\, r^2 V^2 (r) \ln (mr) \nonumber\\
&&-\frac 1 {2\pi} \int_0^\infty dr \,\, V(r)
\int_0^{\infty} d\nu
\frac{\nu} {1+e^{2\pi \nu}}
\ln |\nu^2 -(mr)^2| \nonumber\\
&&-\frac 1 {8\pi} \int_0^\infty dr \,\,
\left[r^2 V^2 (r) -\frac 1 2 V(r)\right]
\int_0^\infty d\nu\, \left( \frac{d}{d\nu} \frac{1}{
1+e^{2\pi \nu}} \right)\, \ln \left| \nu^2 -x^2 \right| \nonumber \\
&&-\frac 1 {8\pi} \int_0^\infty dr \,\, V(r)
\int_0^\infty d\nu\,
\left[ \frac{d}{d\nu} \left( \frac{1}{\nu} \frac{d}{d\nu} \frac{\nu^2}{
1+e^{2\pi \nu}}\right)  \right]\, \ln \left| \nu^2 - x^2 \right|
\label{numasym}\\
&&+\frac 1 {48\pi} \int_0^\infty dr \,\, V(r)
\int_0^\infty d\nu\,
\left[ \frac{d}{d\nu} \left( \frac{1}{\nu} \frac{d}{d\nu}
 \frac{1}{\nu} \frac{d}{d\nu} \frac{\nu^4}{
1+e^{2\pi \nu}}\right)  \right]\, \ln \left| \nu^2 - x^2 \right|,
\nn
\end{eqnarray}
valid for any potential with the above mentioned properties and
very suitable for numerical evaluation.

The remaining task for the analysis of the renormalized
ground state energy
\beq
E_{vac}^{ren} [\Phi ] = E_f  [\Phi ] + E_{as}^{ren} [\Phi ] \nn
\eeq
in the presence
of a spherically symmetric potential is the numerical analysis of $E_f$.
To achieve that, a (as a rule numerical) knowledge of the Jost
function $f_l (ik)$ is necessary.
This will be obtained from a numerical knowledge of the regular solution
at a single point \cite{bord99u}.

Under the assumption that the potential has compact support, let's
say $V(r)=0$ for $r\geq R$, the regular solution may be written as
\begin{eqnarray}
\phi_{l,p}(r) =u_{l,p} (r) \Theta (R-r) +\frac i 2
\left[ f_l(p) \hat h _l^- (pr) - f_l^*(p)
 \hat h _l^+ (pr) \right] \Theta (r-R).\label{gen4.1}
\end{eqnarray}

Assuming continuity of the regular solution and its derivative,
the matching conditions are
\begin{eqnarray}
 u_{l,p} (R) &=& \frac i 2 \left[ f_l(p) \hat h _l^- (pR) - f_l^*(p) \hat h _l^+
 (pR)\right],\nonumber\\
 u_{l,p}' (R) &=& \frac i 2 p \left[ f_l(p)  {\hat h}_l^{- \prime} (pR) -  f_l^*(p)  {\hat h} _l^{+ \prime} (pR) \right].\nonumber
\end{eqnarray}

These conditions provide already the Jost function in terms of the
regular solutions,
\begin{eqnarray}
 f_l(p) = -\frac 1 p \left( p u_{l,p} (R)  {\hat h} _l^{+ \prime} (pR) - u_{l,p}' (R) \hat h _l^+ (pR) \right),\label{gen4.2}
\end{eqnarray}
where we used that the Wronskian determinant of $\hat{h}_l^{\pm}$ is $2i$.

Compared to integral representations of the Jost function \cite{tayl72b},
\begin{eqnarray}
f_l (ik) =1+\int\limits_0^{\infty} dr\,\,r\,V(r) \phi_{l,ik} (r)
   K_{\nu} (kr),\label{gen3.1}
\end{eqnarray}
eq.~(\ref{gen4.2}) has the advantage that the solution is only needed at
the point $r=R$. This simplifies the numerical procedure considerably.

In order to determine a unique solution of the differential equation
(\ref{2.1}), we need to pose an initial value problem. Given that
$\phi_{l,p}(r)$ is the regular solution, we have for $r\to 0$
\begin{eqnarray}
\phi_{l,p}(r) = u_{l,p} (r) \sim \hat j _l (pr)
\sim \frac{\sqrt{\pi}}{\Gamma (l+3/2)} \left( \frac z 2 \right) ^{l+1} .
\nonumber
\end{eqnarray}

The natural ansatz thus is
\begin{eqnarray}
u_{l,p} (r) = \frac{\sqrt{\pi}}{\Gamma (l+3/2)} \left( \frac{pr}
2 \right) ^{l+1} g_{l,p} (r),\nonumber
\end{eqnarray}
with the inital value $g_{l,p} (0) =1$. The differential equation for
$g_{l,p}(r)$ reads
\begin{eqnarray}
\left\{ \frac{d^2}{dr^2} +2\frac{l+1} r \frac d {dr} -V(r) +p^2\right\}
g_{l,p} (r) =0 , \nonumber
\end{eqnarray}
or, going to the needed imaginary $p$-axis and writing it as a first order
differential equation with $(\partial / \partial r) g_{l,ip} (r)
= v_{l,ip } (r)$,
\begin{eqnarray}
\frac d {dr} \left(
\begin{array}{c}
   g_{l,ip} (r) \\
   v_{l,ip} (r)  \end{array} \right) =
\left( \begin{array}{cc}
  0 & 1 \\
 V(r) +p^2 & -\frac 2 r (l+1)
  \end{array} \right)
\left(
\begin{array}{c}
   g_{l,ip} (r) \\
   v_{l,ip} (r)  \end{array} \right).\label{firstord}
\end{eqnarray}

To fix the solution uniquely we only need to fix $v_{l,ip}(0)$. A
power series ansatz for $g_{l,ip} (r)$ about $r=0$ shows that, for
$V(r) = {\cal O} (r^{-1+\epsilon})$, the condition reads
$v_{l,ip}(0) =0$. With this unique solution of (\ref{firstord}),
the Jost function takes the form
\begin{eqnarray}
f_l(ip) = \frac 2 {\Gamma (l+3/2)} \left( \frac{pR} 2 \right)
^{l+3/2} \left\{ g_{l,ip} (R) K_{l+3/2} (pR) +\frac 1 p g'_{l,ip} (R)
 K_{l+1/2} (pR) \right\}.\label{endjost}
\end{eqnarray}

Finally, doing a partial integration and the substitution
$q=\sqrt{k^2-m^2}$ in eq. (\ref{as}), the starting point for the
numerical evaluation used for $E_f$ is
\begin{eqnarray}
E_f = \frac 1 \pi \sum_{l=0}^\infty (l+1/2)
\int_0^\infty dq \,\, \left[ \ln f_l (i\sqrt{q^2+m^2}) -\ln f_l^{asym}
(i\sqrt{q^2+m^2}) \right] .\label{efinfin}
\end{eqnarray}

Now we are prepared to use eqs.~(\ref{numasym}) and
(\ref{efinfin}) for the calculation of $E_{vac}^{ren}$. We choose
a potential with compact support $\Phi (r\geq R) =0$. For the
numerical analysis to come, we have used the dimensionless
quantities \beq \epsilon &=& E_{vac}^{ren} R, \quad \mu =mR, \quad
\rho = \frac r R , \nn\\ V(r) &=& \lambda ' \Phi ^2 (r) = \frac
{\lambda ' } {R^2}  \varphi^2 (\rho ).\nn \eeq

The example we consider here is \cite{bord99u} \beq \varphi (\rho
) = \frac{ 16 a \rho ^2 (1-\rho )^2} {a+(1-2\rho ) ^2 }, \nn \eeq
which is a kind of a spherical wall; the parameter $a$ allows to
vary the shape of the potential (see Fig.~\ref{fig:eb}).

\begin{figure}[!htbp]
  \begin{minipage}[t]{18cm}
\hskip1.5cm
\includegraphics[scale=.5,angle=-90]{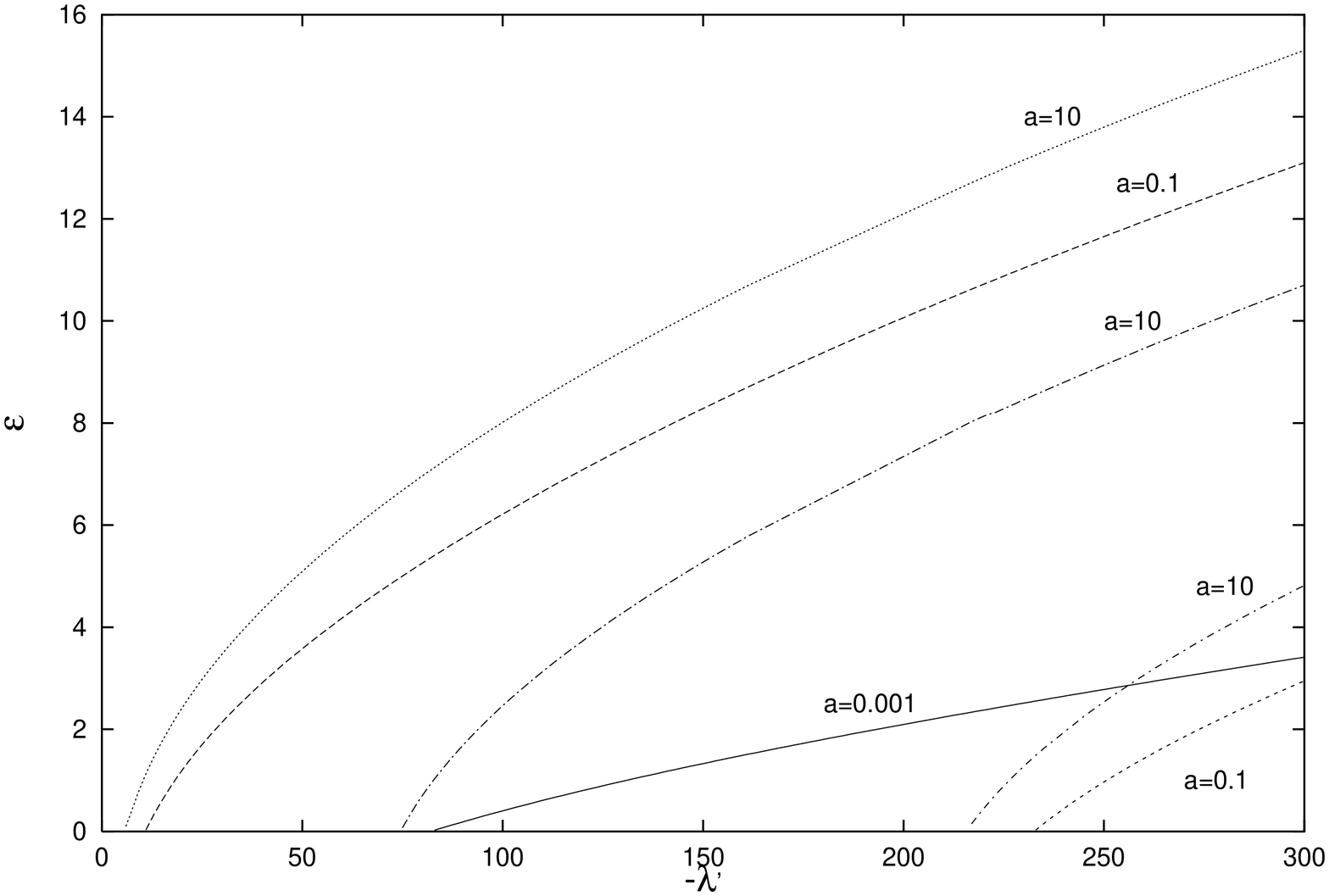}
\end{minipage}
\vskip-8.7cm\hskip2.15cm
  \begin{minipage}[b]{12cm}
\includegraphics[scale=.2,angle=-90]{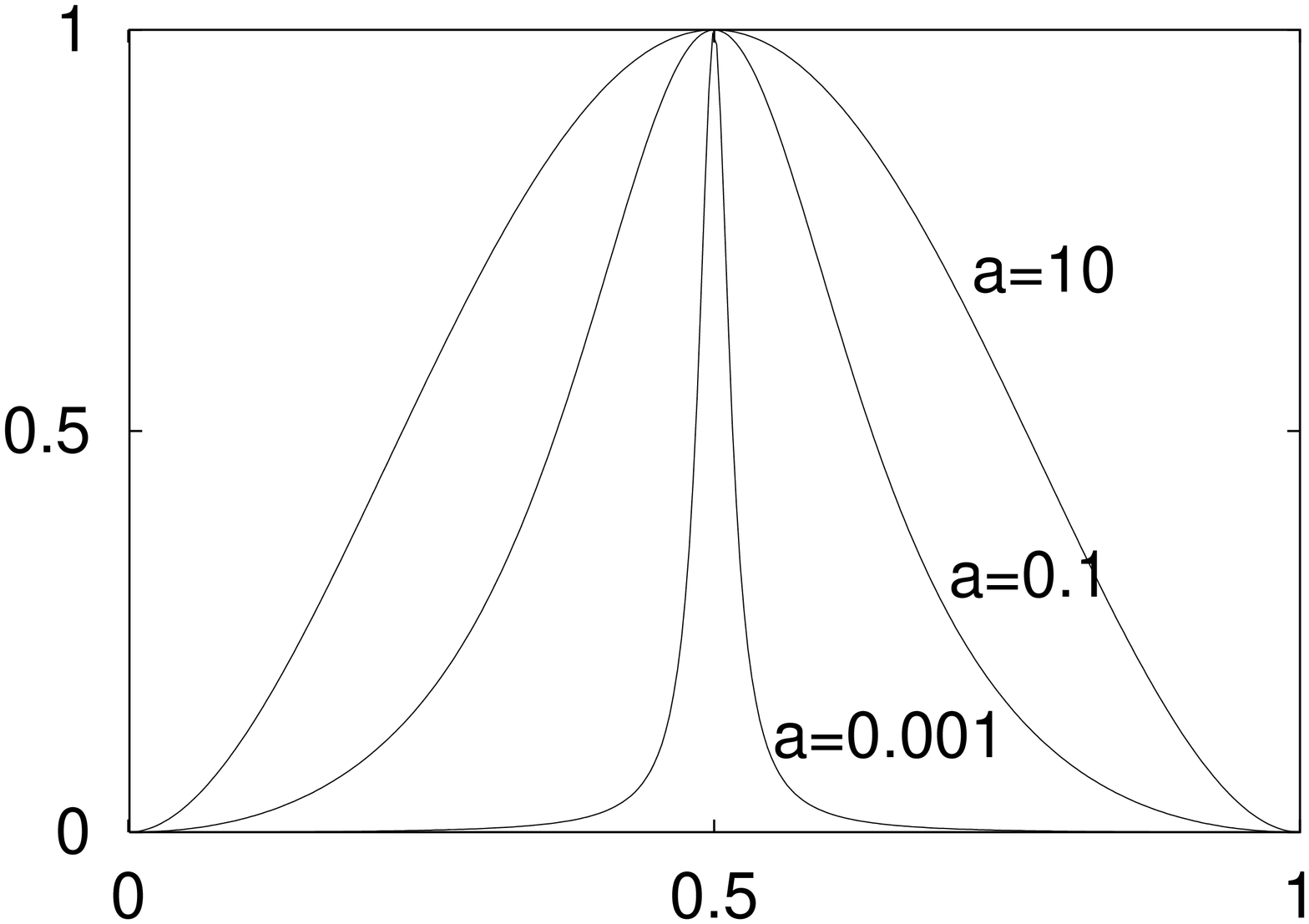}
  \end{minipage}
\vskip5cm
    \caption{Energy of bound states for several shapes of the potential
     and negative $\lambda^{
\prime}$.
(The inset shows  $\varphi(\rho)$.)}
    \label{fig:eb}
\end{figure}

The most determining property of the potential $V(r)$ is the
number and depth of boundstates. The parameter dependence on
$\lambda '$ of the $l=0$ boundstates, determined as zeroes of the
Jost function, are given in Fig.~\ref{fig:eb}. As is clear, number
as well as depth of the boundstates increases with increasing
$(-\lambda ')$, and, as is seen, with increasing $a$.

To show clearly the way this influences the vacuum energy, we consider
the dependence of the energy for fixed $\lambda '$ as a function of the
parameter $a$, Fig.~\ref{fig:eeb}, as well as for fixed $a=1$ as a
function of $\lambda '$, Fig.~\ref{fig:eeb2}.

Let us start with a description of Fig.~\ref{fig:eeb}. For $a$ large
enough, the contribution of the bound state(s) to the vacuum energy
is large enough to overcompensate the scattering state contributions
and the energy is negative. At some "critical value" of $a$, the
energy of the only remaining bound state is so small, that the
positive contributions of the scattering states get the upper hand
and the vacuum energy is positive. Finally, for $a\to 0$ we have
normalized $E_{vac}^{ren} [\Phi =0] =0$, such that at some point,
$E_{vac}^{ren}$ starts to decrease again with decreasing $a$.

\begin{figure}[!htbp]
  \begin{minipage}[t]{18cm}
\hskip1.3cm\includegraphics[scale=.5,angle=-90]{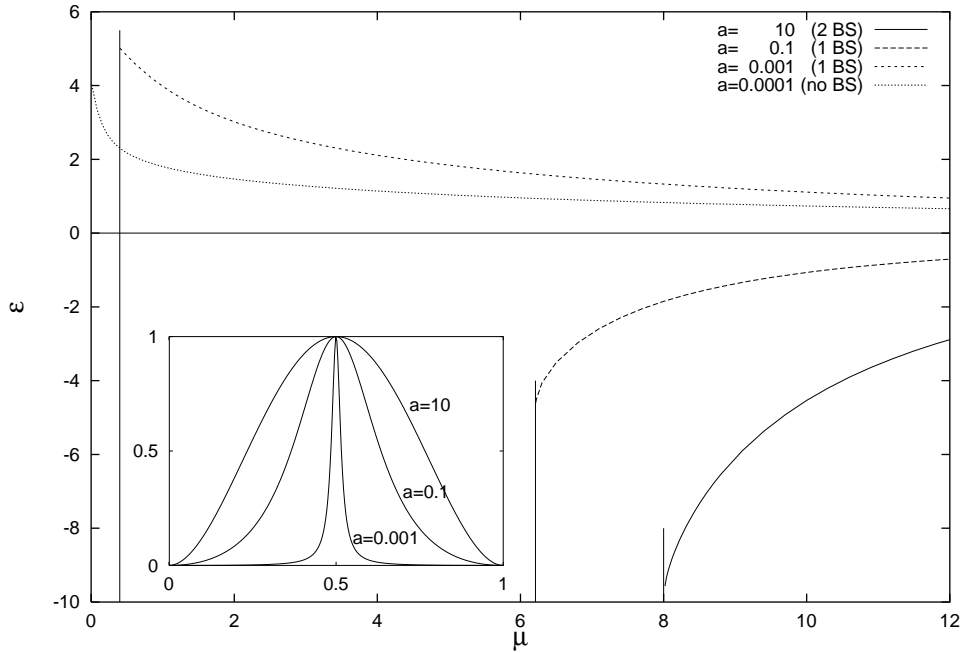}
\end{minipage}
\vskip-4.6cm\hskip2.6cm
  \begin{minipage}[b]{12cm}
\includegraphics[scale=.215,angle=-90]{typeb.eps}
  \end{minipage}
\vskip.5cm
    \caption{Vacuum energy for various shapes of the potential with
          $\lambda^{\prime}=-100$.
      The positions of bound states at the $\mu$ axis are shown as vertical
      lines and the inset shows  $\varphi(\rho)$.}
    \label{fig:eeb}
\end{figure}

The same features are clearly recovered in Fig.~\ref{fig:eeb2}. For large
enough $(-\lambda')$ a negative vacuum energy is obtained. With
decreasing $(-\lambda ' )$ the energy
increases and can be positive or negative depending on the parameter
$\mu$; for fixed value of $R$ this means the mass $m$. Again, at some point
the scattering states get the upper hand, the energy gets positive and
tends to zero as $\lambda '$ tends to zero. For positive $\lambda '$
the energy starts to increase again, being virtually identical for $\lambda =
\pm 2$, and it seems that $E_{vac}^{ren} $ is a monotonically increasing
function of $\lambda ' $ for $\lambda ' \geq 0$.

\begin{figure}[!htbp]
  \begin{minipage}[t]{18cm}
\hskip1.3cm\includegraphics[scale=.5,angle=-90]{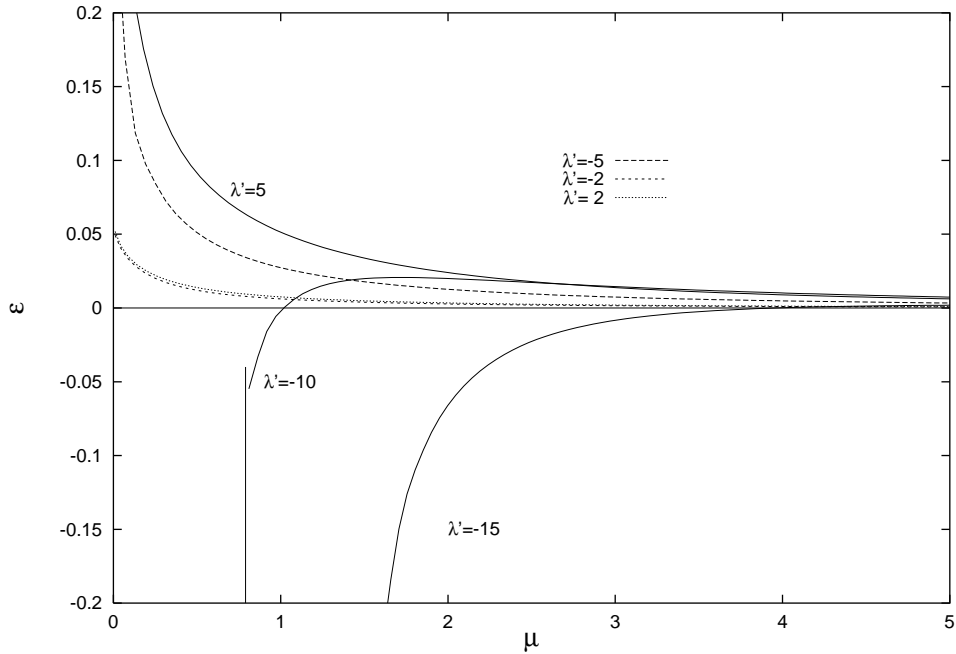}
\end{minipage}
   \caption{Vacuum energy for different magnitudes
$\lambda^{\prime}$
     with equal shape parameter $a=1$.}
    \label{fig:eeb2}
\end{figure}

Exactly the same features are observed if instead of the chosen
spherical wall a lump-like potential concentrated around $r=0$ is
chosen, and it is expected that these features hold for any kind
of potential \cite{bord99u}. In addition, the numerical analysis
seems to indicate that, without bound states, the vacuum energy is
always positive, indicating that scattering states contribute
positively to the vacuum energy (a property we actually used
already in the description of the figures), a result for which no
analytical proof exists in three dimensions.

\section{Conclusions}
In this contribution we have provided methods for the analysis of
zeta functions associated with second order differential
operators, when spherically symmetric external conditions are
present. Basic analytical skills applied are contour and
Mellin-Barnes integral representations and, at a second stage, the
Barnes zeta function. Due to the close connection between zeta
functions and various other spectral functions as the heat-kernel
and determinants, applications to the calculation of heat-kernel
coefficients and ground state energies have been given.

The determination of heat-kernel coefficients is achieved by a
combination of different methods, see Section \ref{sec4}. As a
first step write down the general form of the coefficients in
terms of geometrical invariants. The possible invariants strongly
depend on the boundary conditions considered as has been briefly
described \cite{gilk95b,avra98-15-281}. Next one can consider the
special case calculations of the ball given in Section
\ref{sec3.2} with this general form and fit unknown numerical
constants, as has been explained in detail in Section
\ref{sec4.1b}. Together with information obtained from the product
formula, this gives a very good starting point to apply the
functorial techniques developed in \cite{bran90-15-245}.
Experience shows that, seemingly, any \bc can be successfully
attacked in that way.

Afterwards, we concentrated on the analysis of vacuum energies.
For the case of a massive scalar field inside a spherical shell we
calculated the Casimir energy for the case when the field
satisfies Dirichlet boundary conditions. Contrary to what one
expects from the example of parallel plates, the energy depends
strongly on the mass and, as a function of it, attraction as well
as repulsion is possible. Mathematically, the calculation of the
vacuum energy amounts to an evaluation of the associated zeta
function at the specific point $s=-1/2$. With the help of the
analytical approach provided in Section \ref{sec3.1} this reduces
to a numerical analysis of convergent integrals.

If an external potential is present we have expressed
$E_{vac}^{ren}$ through the associated quantum mechanical
scattering dates, namely the Jost function. This is explicitly
calculated by solving numerically a first order linear
differential equation and by performing numerical integrations.
The vacuum energy behaves in a way one can understand from
physical reasons and the definition introduced by the
normalization condition (\ref{grundnorm}) seems to be very
reasonable.

Let us emphasize that the external potential can be a solution of
some classical nonlinear equation. But to apply the method one
needs just any spherically symmetric potential of otherwise
whatever shape to come immediately to a determination of
$E_{vac}^{ren}$. This is hoped to find further applications in the
context of the fields mentioned briefly in the Introduction.

\vspace{0.5cm}
{\bf Acknowledgement:} I am indebted to the institute of physics of the
university of La Plata for the invitation to the conference and for the
very kind hospitality during my stay
(partially supported by ANPCyT of Argentina, under grant PICT 00039).
Several discussions with the members
of the theory group, especially Mariel Santangelo and Horacio Falomir,
have been influential on the presentation as well as the content of the
contribution. Finally, let me thank Michael Bordag, Stuart Dowker and
Emilio Elizalde for the pleasant and fruitful scientific collaboration 
on the subjects presented. The work is supported by the DFG under 
contract number Bo 1112/4-2.

\renewcommand{\theequation}{\Alph{section}.\arabic{equation}}
\begin{appendix}

\section{Representations for the asymptotic contributions}
\label{anhda}
\setcounter{equation}{0}
In this Appendix we derive explicit representations of the
asymptotic contributions in the Casimir and ground state energies for
massive fields. Let us start with $A_{-1} (s)$ for the massive
scalar field, eq.~(\ref{eq2.14}), which, a little bit more explicit
reads
\begin{eqnarray}
A_{-1} (s) = 2\frac{\sin (\pi s)}{\pi} \sum_{l=0}^{\infty} \nu^2
\int\limits_{ma/\nu}^{\infty} \left[\left(\frac{x\nu}a\right)^2-m^2
\right]^{-s} \frac{\sqrt{1+x^2}-1} x, \label{anhdaa1}
\end{eqnarray}
of which we need the analytical continuation to $s=-1/2$.
With the substitution $t=(x\nu /a)^2 -m^2$, this expression results into
the following one
\begin{eqnarray}
A_{-1} (s)
&=& \frac{\sin (\pi s)}{\pi} \sum_{l=0}^{\infty} \nu \int\limits _0 ^{\infty}
dt\,\,\frac{t^{-s}}{t+m^2}\left\{
\sqrt{\nu^2 +a^2 (t+m^2)} -\nu\right\}\nonumber\\
&=&-\frac 1 {2\sqrt{\pi}}
 \frac{\sin (\pi s)}{\pi} \sum_{l=0}^{\infty} \nu \int\limits _0 ^{\infty}
dt\,\,t^{-s} \int\limits _0^\infty
d\alpha \,\, e^{-\alpha (t+m^2)}\label{anhdaa2}\\
& &\times\int\limits_0^{\infty}d\beta \,\,\beta ^{-3/2} \left\{
e^{-\beta (\nu^2 +a^2 [t+m^2] ) } -e^{-\beta \nu^2}\right\},\nonumber
\end{eqnarray}
where the Mellin integral representation for the single factors has been
used. As we see,
the $\beta$-integral is well defined. Introducing a regularization parameter
$\delta$, $A_{-1}(s)$ can then be written as
\begin{eqnarray}
A_{-1} (s) =\lim_{\delta \to 0} \left[ A_{-1}^1 (s,\delta )
+A_{-1} ^2 (s,\delta ) \right], \label{anhdaa3}
\end{eqnarray}
with
\begin{eqnarray}
A_{-1}^1 (s,\delta )
 &=&-\frac 1 {2\sqrt{\pi}}
 \frac{\sin (\pi s)}{\pi} \sum_{l=0}^{\infty} \nu
\int\limits_0^\infty  d\alpha \,\,
 e^{-\alpha m^2}
\int\limits_0^{\infty}d\beta \,\,\beta ^{-3/2+\delta}e^{-\beta (\nu^2 +a^2
m^2 ) }
\int\limits _0 ^{\infty}
dt\,\,t^{-s}e^{-t(\alpha +\beta a^2)}\nonumber
\end{eqnarray}
and
\begin{eqnarray}
A_{-1} ^2 (s,\delta ) =    \frac 1 {2\sqrt{\pi}} \Gamma (1-s)
\frac{\sin (\pi s)}{\pi}    \sum_{l=0}^{\infty} \nu
\int\limits_0^\infty  d\alpha \,\,
e^{-\alpha m^2 }   \alpha^{s-1}\int\limits_0^{\infty}d\beta \,\,
\beta ^{-3/2+\delta}e^{-\beta \nu^2}.
\nonumber
\end{eqnarray}

Let us proceed with the remaining pieces. In $A_{-1}^1 (s,\delta
)$ two of the integrals can be done, yielding
\begin{eqnarray}
A_{-1}^1 (s,\delta )
&=& -\frac{a^{1-2\delta}}{2\sqrt{\pi}\Gamma (s)} \Gamma (s+\delta -1/2)
\times\label{anhdaa4}\\
& &
\sum_{l=0}^{\infty} \nu \int\limits  _0 ^{\infty}
dy\,\, y^{\delta -3/2} \left[ m^2 +y \left(\frac{\nu} a\right)^2
\right]^{1/2-s-\delta}. \nonumber
\end{eqnarray}

For $A_{-1} ^2 (s,\delta )$, one gets
\begin{eqnarray}
A_{-1} ^2 (s,\delta )&=& \frac{m^{-2s}}{2\sqrt{\pi}}\Gamma (\delta -1/2)
\sum_{l=0}^{\infty} \nu^{2-2\delta} \nonumber\\
&=& \frac{a^{1-2\delta}}{2\sqrt{\pi}\Gamma (s)}
\Gamma (s+\delta -1/2)
\sum_{l=0}^{\infty} \nu \int\limits  _0 ^{\infty}
dx\,\, x^{s-1        }
\left[ m^2 x +
\left(\frac{\nu} a\right)^2 \right]^{1/2-s-\delta}.
\label{anhdaa5}
\end{eqnarray}

Adding up (\ref{anhdaa4}) and (\ref{anhdaa5}) yields
\begin{eqnarray}
A_{-1} (s) &=& \frac a {2\sqrt{\pi}\Gamma (s)}
\Gamma (s -1/2) \sum_{l=0}^{\infty} \nu \int\limits  _0 ^{1}
dx\,\, x^{s-1        }
\left[ m^2 x +
\left(\frac{\nu} a\right)^2 \right]^{1/2-s} \nn\\
&=& \frac{a^{2s}}{2\sqrt{\pi}} \frac{\G (s-1/2)}{\G (s+1)}
      \sum_{l=0}^\infty \nu \left\{ \frac 1
     {\left[\nu^2 +(ma)^2\right] ^{s-1/2}} \right.\nn\\
& &\left.\hspace{1cm}\quad+\left( s-\frac 1 2 \right) (ma)^2
           \int_0^1 dx \,\, \frac{x^s}{(\nu^2 +(ma)^2 x)^{s+1/2}} \right\},
\label{anhdaa6}
\eeq
where in the last step a partial integration has been performed such that
the $x$-integral is well behaved at $s=-1/2$.

This is a form suited for the treatment of the angular momentum sum,
which is performed using
\begin{eqnarray}
\sum _{\nu=1/2} ^{\infty} f( \nu ) =\int_0^\infty d\nu \,\, f(\nu ) -i
\int_0^{\infty}d\nu \,\, \frac{f(i\nu +\epsilon )- f(-i\nu +\epsilon )}
{1+e^{2\pi \nu}},\label{anhdaa7}
\end{eqnarray}
where $\epsilon\to 0$ is understood and appropriate analytic properties
of the function $f(\nu)$ are assumed. This is often used
in finite temperature field theory and is equivalent there to switching
from imaginary time to real time \cite{kapu89b}.
When expanding the function $f(\nu)$ in a Taylor
series, one arrives at the well known Euler-Maclaurin summation formula
(a thorough treatment of the Euler-Maclaurin summation formula can be found
in Ref.~\cite{wong89b}).

For $A_{-1}(s)$ the relevant application of eq.~(\ref{anhdaa7}) is
\beq
\sum_{\nu =1/2,3/2,...}^{\infty} \nu^{2n+1} \left(1+\left(\frac{\nu} x
\right) ^2 \right) ^{-s} &=& \frac 1 2 \frac {n! \Gamma (s-n-1)}{\Gamma (s)}
x^{2n+2} \label{anhdaa8}\\
& &\hspace{-25mm} +(-1)^n 2 \int\limits_0^x d\nu \,\, \frac{\nu^{2n+1} }
{1+e^{2\pi\nu}  } \left( 1-\left( \frac{\nu} x \right)^2 \right) ^{-s} \nn\\
& & \hspace{-25mm} +(-1)^n 2\cos (\pi s)
 \int\limits_x^{\infty} d\nu \,\, \frac{\nu^{2n+1} }
{1+e^{2\pi\nu}  } \left( \left( \frac{\nu} x \right)^2 -1
\right) ^{-s}. \nn
\eeq

Together with the integral \beq \int_0^\infty \frac
{\nu^n}{1+e^{2\pi \nu}} = \frac{n!}{(2\pi)^{n+1}}
         \eta (n+1),\nn
\eeq where $\eta (s)$ is the eta-function, \beq \eta (s) =
\sum_{k=1}^\infty \frac{(-1)^{k+1}}{k^s} ,\nn \eeq the following
final form of $A_{-1} (s)$ is obtained, \beq A_{-1} (-1/2 +s)
&=&\left(\frac 1 s +\ln a^2\right) \left(\frac 7 {1920\pi a}
+\frac{m^2a}{48\pi} -\frac{m^4a^3}{24\pi}\right)\nonumber\\ &
&+\ln4\left(\frac 7 {1920\pi a} +\frac{m^2a}{48\pi}
-\frac{m^4a^3}{24\pi}\right)\label{anhdaa}\\ & &+\frac 7 {1920\pi
a} -\frac{m^2a}{48\pi}+\frac{m^4a^3}{48\pi} \left(1+4\ln
(ma)\right)\nonumber\\ & &-\frac 1 {\pi a}
\int\limits_0^{\infty}d\nu\,\, \frac{\nu}{ 1+e^{2\pi\nu}}
(\nu^2-m^2a^2 )\ln|\nu^2-m^2a^2 |\nonumber\\ & &-\frac{2m^2a}{\pi}
\int\limits_0^{\infty}d\nu\,\, \frac{\nu}{ 1+e^{2\pi\nu}}\left(\ln
|\nu^2-m^2a^2 |
+\frac{\nu}{ma}\ln\left|\frac{ma+\nu}{ma-\nu}\right|\right).\nonumber
\end{eqnarray}

In the same way, for
\begin{eqnarray}
A_0 (s) &=& -\frac{m^{-2s}} 2 \sum_{l=0}^{\infty} \nu \left[
1+\left( \frac{\nu}{ma}\right)^2 \right]^{-2s}, \nn \eeq using
(\ref{anhdaa8}), it follows
\begin{eqnarray}
A_0(s)&=&\frac 1 6 a^2m^3 -m\int\limits_0^{ma}d\nu\,\,
 \frac{\nu}{ 1+e^{2\pi\nu}}\sqrt{1-\left(\frac{\nu}{ma}\right)^2}.
\label{anhdab}
\end{eqnarray}

For the remaining asymptotic contributions, the central spectral
function is, see eq.~(\ref{fab}), \beq f(s;c;b) =\sum_{\nu =
1/2,3/2,\ldots} ^{\infty} \nu ^c \left( 1+ \left( \frac{\nu} {ma }
\right ) ^2 \right) ^{-s-b}.\nn \eeq

To deal with all values of $c$ and $b$ needed, in addition to
eq.~(\ref{anhdaa8}), we need \beq \sum_{\nu =1/2,3/2,...}^{\infty}
\nu^{2n} \left(1+\left(\frac{\nu} x \right) ^2 \right) ^{-s} &=&
\frac 1 2 \frac { \Gamma (n+1/2) \Gamma (s-n-1/2)}{\Gamma (s)}
x^{2n+1} \label{anhdaa9}\\ & &\hspace{-25mm} -(-1)^n 2\sin (\pi s)
 \int\limits_x^{\infty} d\nu \,\, \frac{\nu^{2n} }
{1+e^{2\pi\nu}  } \left( \left( \frac{\nu} x \right)^2 -1
\right) ^{-s}. \nn
\eeq

Using partial integrations one can obtain representations valid
for values of $s$ needed for the asymptotic contributions. To
simplify notation, in the following we shall use $x$ for $(ma)$
and $f(c;b) = f(-1/2;c;b)$. A full list of ingredients needed is
the following: \beq && f(1;1/2) = -\frac 1 2 x^2 +\frac 1 {24}, \nn
\\ &&\frac d {ds} \left|_{s=-1/2} \right. f(s;1;1/2) = -\frac 1 2
x^2 -2\int_0^{\infty} d\nu \frac{\nu} {1+e^{2\pi \nu}} \ln
\left|1-\left(\frac{\nu}{x}\right)^2 \right| ,\nn\\
&&f(1;3/2)=\frac{x^2}{2(s+1/2)} + x^2 \ln x + x^2 \int_0^\infty
d\nu\, \left( \frac{d}{d\nu} \frac{1}{ 1+e^{2\pi \nu}} \right)\,
\ln \left| \nu^2 -x^2 \right|, \nn \\ &&f(1;1)=2x^2 \int_0^x
d\nu\, \left( \frac{d}{d\nu} \frac{1}{ 1+e^{2\pi \nu}} \right)\,
\left[ 1 - \left( \frac{\nu}{x} \right)^2
 \right]^{1/2}, \nn \\
&&f(1;2)=-2x^2 \int_0^x d\nu\, \left( \frac{d}{d\nu} \frac{1}{
1+e^{2\pi \nu}} \right)\, \left| 1 - \left( \frac{\nu}{x} \right)^2
 \right|^{-1/2}, \nn \\
&&f(3;2)=\frac 5 2 x^4 +2x^4 \int_0^x d\nu \,
\left[ \frac{d}{d\nu} \left( \frac{1}{\nu} \frac{d}{d\nu} \frac{\nu^2}{
1+e^{2\pi \nu}}\right)  \right]\,  \left[ 1 - \left( \frac{\nu}{x} \right)^2
 \right]^{1/2}, \nn \\
&&f(3;5/2)=\frac{x^4}{2(s+1/2)}+(\ln x -1/2) x^4 +\frac{x^4}{2}
 \int_0^\infty d\nu\,
\left[ \frac{d}{d\nu} \left( \frac{1}{\nu} \frac{d}{d\nu} \frac{\nu^2}{
1+e^{2\pi \nu}}\right)  \right]\, \ln \left| \nu^2 - x^2 \right|, \nn \\
&&f(3;3)=  - \frac{2}{3} x^4  \int_0^x d\nu\,
\left[ \frac{d}{d\nu} \left( \frac{1}{\nu} \frac{d}{d\nu} \frac{\nu^2}{
1+e^{2\pi \nu}}\right)  \right]\,  \left[ 1 - \left( \frac{\nu}{x} \right)^2
 \right]^{-1/2}, \nn \\
&&f(5;7/2)=\frac{x^6}{2(s+1/2)}+(\ln x -3/4) x^6\nn \\
&& \hspace{3cm} +\frac{x^6}{8}
 \int_0^\infty d\nu\,
\left[ \frac{d}{d\nu} \left( \frac{1}{\nu} \frac{d}{d\nu}
 \frac{1}{\nu} \frac{d}{d\nu} \frac{\nu^4}{
1+e^{2\pi \nu}}\right)  \right]\, \ln \left| \nu^2 - x^2 \right|, \nn \\
&&f(7;9/2)=\frac{x^8}{2(s+1/2)}+(\ln x -11/12) x^8 \nn \\ &&
\hspace{3cm}  +\frac{x^8}{48}
 \int_0^\infty d\nu\,
\left[ \frac{d}{d\nu} \left( \frac{1}{\nu} \frac{d}{d\nu}
 \frac{1}{\nu} \frac{d}{d\nu} \frac{1}{\nu} \frac{d}{d\nu} \frac{\nu^6}{
1+e^{2\pi \nu}}\right)  \right]\, \ln \left| \nu^2 - x^2 \right|.
\label{eqe12}
\eeq

\end{appendix}


\end{document}